\documentclass[showpacs,aps,pre]{revtex4}
\usepackage{amsmath}
\usepackage{graphicx}
\include{epsf}

\def\(({\left(}
\def\)){\right)}                       
\def\[[{\left[}
\def\]]{\right]}    

\newcommand{\<}{\langle}
\renewcommand{\>}{\rangle}
\newcommand{\be}{\begin{equation}}
\newcommand{\ee}{\end{equation}}
\newcommand{\bea}{\begin{eqnarray}}
\newcommand{\eea}{\end{eqnarray}}

\begin{document}

\title{Threshold values, stability analysis and high-$q$ asymptotics
  for the coloring problem on random graphs}

\author{Florent Krz\c{a}ka{\l}a} \affiliation{Dipartimento di Fisica,
INFM and SMC, Universit\`a di Roma ``La Sapienza'', P.~A.~Moro 2,
I-00185 Roma, Italy}

\author{Andrea Pagnani} \affiliation{Institute for Scientific
Interchange (ISI), Viale Settimio Severo 65, I-10133 Torino, Italy}
 
\author{Martin Weigt} \affiliation{Institute for Theoretical Physics,
University of G\"ottingen, Tammannstr. 1, D-37077 G\"ottingen, Germany}

\begin{abstract}
  We consider the problem of coloring Erd\"os-R\'enyi and regular
  random graphs of finite connectivity using $q$ colors. It has been
  studied so far using the cavity approach within the so-called
  one-step replica symmetry breaking (1RSB) ansatz. We derive a
  general criterion for the validity of this ansatz and, applying it
  to the ground state, we provide evidence that the 1RSB solution
  gives {\it exact} threshold values $c_q$ for the $q$-COL/UNCOL phase
  transition. We also study the asymptotic thresholds for $ q \gg 1$
  finding $c_q = 2q\log{q}-\log{q}-1+o(1)$ in perfect agreement with
  rigorous mathematical bounds, as well as the nature of excited
  states, and give a global phase diagram of the problem.
\end{abstract}


\maketitle
\section{Introduction}

The graph coloring problem (COL) has been studied both in
combinatorics \cite{GaJo} and in statistical physics \cite{WU}. Given
a graph, or a lattice, and given a number $q$ of available colors, the
problem consists in assigning a color to each vertex such that no edge
has two equally colored end vertices. For a given graph, one quantity
of interest is thereby the minimal number of colors needed, i.e. the
so-called {\it chromatic number}.

In this paper we are going to consider COL as applied to random graphs
of fluctuating as well as of fixed connectivity. In fact, determining
their chromatic number is one of the most fundamental open problems in
random-graph theory \cite{Bo}. It has attracted considerable interest
also within the theoretical computer-science literature: COL is one of
the basic NP-hard problems which form the very core of complexity
theory \cite{GaJo}. Defined on random graphs, the problem shows
interesting phase transitions at the so-called $q$-COL/UNCOL
thresholds $c_q$: Graphs of average connectivity $c<c_q$ owe proper
$q$-colorings with high probability (approaching one for graph size
$N\to\infty$), whereas graphs of higher connectivity require more than
$q$ colors.  This transition is connected to a pronounced peak in the
numerical resolution time, i.e.~in the time needed to either construct
a $q$-coloring or to prove its non-existence. The hardest to solve
problems are typically situated close to the phase boundary.

One of the first important mathematical results for $q$-COL on
Erd\"os-R\'enyi random graphs \cite{Erdos_Renyi} of average connectivity
$c$ was obtained by {\L}uczak more than one decade ago \cite{Lu}. He
showed in particular that, for a random graph of given finite average
connectivity $c$, the chromatic number takes one out of only two
possible consecutive values with high probability. Even if he was not
yet able to determine these values, he showed that $q$ colors are {\it
  not} sufficient for almost all graphs with $c \geq 2q\ln q-\ln q - 1
+ o(1)$. Rephrased in terms of the $q$-COL/UNCOL threshold $c_q$, he
thus proved the upper bound $c_q \leq 2q\ln q-\ln q - 1 + o(1)$. Very
recently Achlioptas and Naor \cite{Ach_Nao} put a rigorous lower bound
on the threshold using the second moment method.  They showed that
$c_q \geq 2q\ln q - 2\ln q + o(1)$, but their method in fact leads to
an even better conjectured lower bound $c_q \geq 2q\ln q - \ln q - 2 +
o(1)$ \cite{Ac_note} which differs just by one from {\L}uczak's upper
bound.  So, up to these small intervals, the exact and unique value
of the chromatic number is known by now.  Inside these intervals,
even more powerful methods are needed to determine the chromatic number
and thus the COL/UNCOL threshold.

If one considers, on the other hand, the performance of linear-time
algorithms, $q$-colorings can be easily constructed up to connectivity
$c\simeq q\ln q$, i.e.~only in roughly the first half of the colorable
phase. It is simple to design algorithms working up to
$c=(1-\varepsilon) q\ln q$, for any $\varepsilon>0$, whereas no
linear-time algorithm is known which works also for connectivity
$c=(1+\varepsilon) q\ln q$ \cite{ShUp,FeVe,AcMo}. The very existence
of linear algorithms working also beyond this point is considered as
another major open question \cite{FrMcDi} within the field.

Recently, the problem has been reconsidered using tools borrowed from
statistical mechanics of disordered systems
\cite{Coloring1,Coloring2}. In this way both questions, i.e.~the
location of the $q$-COL/UNCOL threshold and the reason for the failure
of linear-time algorithms well before this threshold, have further
approached an answer, though not on completely rigorous grounds.
Within the 1RSB approach, the $q$-COL/UNCOL transition $c_q$ can be
determined for an arbitrary number of colors $q$. Moreover, the 1RSB
approach predicts a connectivity region $c_d<c<c_q$ inside the
colorable phase, where solutions are non-trivially organized in
clusters, an exponential number of metastable states and large
energetic barriers exist. This clustering phenomenon -- intuitively --
causes local algorithms to get stuck, see \cite{MoRi2} for recent
results.

All these results were derived with the cavity method in 1RSB
approximation \cite{MePa} which makes strong hypotheses on the phase
space structure of the problem. The computation is further on heuristic
in the sense that assumptions are only checked self-consistently, but
its predictions are confirmed by independent numerical tests. Very
recently a number of papers have put under scrutiny the clustering
hypothesis in glass models on the Bethe lattice~\cite{Rivoire}, in
some combinatorial optimization problems like random $K$-XORSAT and
$K$-SAT \cite{MoRi,MonRicci,MerMezZec} as well as in polymer
problems~\cite{MMM}. They analyze the possibility of more complex
patterns of clusterization due to local instabilities. 

In this work we investigate these instabilities for the coloring
problem. We thereby show that a small part of the 1RSB solution of
\cite{Coloring1, Coloring2} close to the onset of the clustered phase
turns out to be unstable.  Interestingly enough, however, the
$q$-COL/UNCOL transition is in the stable region at any $q$, and thus
the 1RSB threshold results are expected to be exact. We will also
analyze the stability of the coloring problem on fixed connectivity
random graphs, which is somehow easier to deal with analytically.

The outline of the paper is as follows: Sec.~\ref{Sec:SP} properly
defines the problem under investigation, and reviews the 1RSB
approach.  In Sec.~\ref{Sec:Stab} we set up the general formalism for
the stability analysis. The most relevant consequences of this
approach for a small number of available colors are then presented in
Sec.~\ref{Sec:TypeI} for the TYPE-I instability, and in
Sec.~\ref{Sec:TypeII} for the one of TYPE-II.  Sec.~\ref{Sec:High-q}
is devoted to the high $q$ analysis of the model while, in
Sec.~\ref{Sec:Finite}, we finally consider the problem at finite
energies. Conclusions and perspectives are drawn in
Sec.~\ref{Sec:Concl}.

\section{The model and its 1RSB solution}
\label{Sec:SP}

\subsection{The graph coloring problem}
\label{sec:colintro}

Let us start with a proper definition of the problem.  We consider a graph
$G = ({\cal V,E})$ defined by its vertices ${\cal V}=\{1,...,N\}$ and
undirected edges $(i,j)\in {\cal E}$ which connect pairs of vertices $i,j\in
{\cal V}$. A {\it graph $q$-coloring} is a mapping $\sigma:\ {\cal V}\to
\{1,...,q\}$ which assigns colors $1,...,q$ to all vertices, such that
no edges are monochromatic. For all edges $(i,j)\in {\cal E}$ we have
therefore $\sigma_i \neq \sigma_j$.

Within the statistical-mechanics approach, a Hamiltonian is assigned
to this problem such that all $q$-colorings are found as ground
states. For any color assignment, i.e. $\{\sigma_i\} \in
\{1,2,\dots,q\}$ for all $i\in {\cal V}$, we therefore define
\begin{equation}
\label{eq_hamilton}
H_G = \sum_{(i,j) \in {\cal E}} \delta(\sigma_i,\sigma_j)
\end{equation}
with $\delta(\cdot,\cdot)$ denoting the Kronecker symbol. This
Hamiltonian counts the number of monochromatically colored edges, a
proper coloring of the graph thus has zero energy. In a physicist's
language, the Hamiltonian describes an anti-ferromagnetic $q$-state
Potts model on the graph $G$.

The aim within the statistical mechanics approach is to study the
ground state properties of this model: If the ground state energy
equals zero, the graph is colorable. The ground-state entropy
determines the number of colorings, and the order parameter, see
below, characterizes the statistical properties of the ensemble of all
solutions. If, on the other hand, the ground-state energy becomes
positive, we know that there are no proper colorings. The graph is
uncolorable with $q$ colors.

\subsection{Erd\"os-R\'enyi random graphs and regular random graphs}
\label{sec:graphs}

We consider the graph coloring problem on two different random-graph
ensembles.

The first one is the ensemble ${\cal G}(N, c/(N-1))$ first introduced
by Erd\"os and R\'enyi in the late 1950s \cite{Erdos_Renyi}. A graph
from this ensemble consists of $N$ vertices $j=1,...,N$. Between each
pair $i,j$ of vertices, with $i<j$, an undirected edge is drawn
randomly and independently with probability $c/(N-1)$. The vertices
remains unconnected by a direct edge with probability $1-c/(N-1)$.

Here we are mainly interested in the thermodynamic limit $N\to\infty$,
i.e. we describe large graphs of finite $c$. The average vertex
degree, which equals the expected number of edges incident to an
arbitrary vertex, is easily calculated as $(N-1)\cdot c/(N-1) = c$,
and it remains finite in the large-$N$ limit. There are, however,
degree fluctuations for every finite $c$. In fact, in the
thermodynamic limit, the probability that a randomly selected vertex
has degree $d$, is given by the Poissonian distribution
\begin{equation}
  \label{eq:poisson_deg}
  p_d = e^{-c} \frac {c^d}{d!}
\end{equation}
of mean $c$. Another crucial point for our analysis is that, for
finite $c$, the number of triangles or other short loops in the graph
remains finite in the large-$N$ limit. This means that the graph is
almost everywhere {\it locally tree-like}, i.e. on finite length scales
it looks like a tree. For $c>1$, there exists an extensive number of
loops. These have, however, length ${\cal O}(\ln N)$, and become
infinitely long for $N\to\infty$.
 
The second ensemble is denoted by ${\cal G}_{c}(N)$, and contains
all {\it $c$-regular} graphs of $N$ vertices, where $c$ has to be a
positive integer in this case. A graph is called $c$-regular if and
only if {\it all} vertices have the same degree $c$, i.e. here we
have 
\begin{equation}
  \label{eq:constant_deg}
  p_d = \delta (d,c) \ .
\end{equation}
A random regular graph is one randomly selected element of this
ensemble. This guarantees again that the graph becomes locally
tree-like. Note that due to the constant vertex degree, these graphs
look locally homogeneous, on finite length scales they do not show any
disorder. The random character of regular graphs enters only via
the long loops which are again of length ${\cal O}(\ln N)$. In the
statistical physics literature, in particular in the theory of
disordered and glassy systems, random graphs are considered as one
valid definition of a Bethe lattice.

In a slightly more general context, both random graph ensembles
defined before can be embedded into the ensemble of random graphs with
given degree distribution \cite{MolRee,NewStrWat}. For these graphs,
only $p_d$ is defined, and each graph having the desired degree
distribution is considered to be equiprobable. In this sense, all
results formulated below can be directly generalized to arbitrary
degree distributions.  Sometimes the formulation will even be given in
this context, but the concrete analysis will be restricted to
Poissonian and regular graphs.

For this generalized ensemble, we still introduce the probability $r_d$
that an arbitrary end-vertex of a randomly selected edge has {\it
  excess degree} $d$, i.e. it is contained in $d$ supplementary edges,
and its total vertex degree is $d+1$. Given the degree distribution,
this probability results as
\begin{equation}
  \label{eq:neighbors}
  r_d = \frac {(d+1) p_{d+1}} c \ .
\end{equation}
Plugging in our special cases, we see that this distribution remains
Poissonian for Erd\"os-R\'enyi graphs, whereas the excess degrees
equal constantly $c-1$ in the case of regular graphs.

\subsection{Survey propagation equations for $q$-coloring}
\label{sec:sp_eq}

The cavity equations for finite-connectivity systems in the one-step
replica-symmetry broken approximation have been originally derived
in~\cite{MePa,Cavity}. Their single sample version is usually called
survey propagation (SP) and has been introduced in~\cite{MePaZe,MeZe}
in the case of random 3-SAT. Here, in order to fix the notation, we
will briefly recall the SP equations on the 1RSB level for $q$-COL
closely following~\cite{Coloring1,Coloring2}. Note, however, that a
complete explanation of all technical details is not the scope of this
paper, for a detailed presentation please see therefore the original
publications~\cite{Coloring1,Coloring2}.

The zero-temperature properties of the system (or local minima of
$H_G$) can be completely characterized by the edge-dependent
probability distributions ($(i,j)\in{\cal E}$):
\begin{equation}
\label{Qu}
Q_{i\rightarrow j}(\vec u) = \eta_{i\rightarrow j}^0\delta(\vec u) +
\sum_{\tau=1}^q \eta_{i\rightarrow j}^\tau \delta(\vec u + \vec
e_\tau)
\end{equation} 
where the vectors $\{\vec e_1, \cdots, \vec e_q\}$ form the usual
$q$-dimensional canonical Euclidean base set, with components
$e^\sigma_\tau = \delta(\tau, \sigma )$. The $\eta_{i\rightarrow j}$s
are positively defined probabilities, normalization implies
$\sum_{\tau=0}^q \eta_{i\rightarrow j}^\tau = 1$. The distribution
$Q_{i\rightarrow j}(\vec u)$ is called a {\it survey}, and it
describes the probability that, in a suitably chosen metastable state,
or local minimum of $H_G$, a {\it warning} $\vec u$ is send from
vertex $i$ via the edge $(i,j)$ to vertex $j$. Possible warnings are
the vectors $-\vec e_\tau$ which include a warning that assigning
color $\tau$ to vertex $i$ will cause an energy increase, and the zero
message $\vec u = (0,...,0)$. In the following, warnings will
frequently denoted simply by their indices $\tau=0,...,q$.

These distributions are self-consistently determined via the SP
equations 
\begin{eqnarray}
\label{eq_surv_finiteyI}
P_{i\rightarrow j}(\vec h)&=& C_{i\rightarrow j} \int \Big[ \prod_{k
\in V(i) \setminus j} d^q \vec u_k Q_{k\rightarrow i}(\vec u_k)\Big]
\delta \left(\vec h-\sum_{k\in V(i)\setminus j} \vec u _k\right) \exp
\left\{ y\ \omega\left(\sum_{k\in V(i)\setminus j} \vec u_k\right)
\right\}\\
\label{eq_surv_finiteyII}
Q_{i\rightarrow j}(\vec{u}) &=&\int d^q \vec h\ 
P_{i\rightarrow  j}(\vec h)
\delta\left(\vec u- \hat u  ( \vec h ) \right)\ .
\end{eqnarray}
There the $C_{i\rightarrow j}$ are normalization constants, the set
$V(i)\subset {\cal E}$ contains all neighbors of vertex $i$,
and the functions $\hat u$ and $\omega$ are defined as:
\begin{eqnarray}
\label{eq:u}
\omega(\vec h) &=& - \min ( -h^1,..., -h^q )\nonumber\\ 
\hat u^\tau(\vec h) &=& \omega(\vec h-\vec e_\tau) - \omega(\vec
h)\ . 
\end{eqnarray}
These equations have a very nice interpretation: A site $i$ receives
an incoming field $\vec h$ as a sum of all but one incoming warnings.
This field has either zero or negative entries, reflecting the
anti-ferromagnetic character of the interaction. The maximal field
components determine the colors of minimal energy if assigned to $i$.
If this maximal field component is unique, a non-trivial message ``Do
not take this unique color!'' is sent from vertex $i$ to $j$ via the
last link. If the maximal field component is degenerate, the zero
message is sent via link $i\to j$.

Note the appearance of the reweighting parameter $y>0$, which allows
to scan metastable states of different energies. It acts similar to
the inverse temperature in the usual Boltzmann weight: Different $y$
concentrate the measure on different energy levels, and the limit
$y\to \infty$ corresponds to zero-energy ground states.

The corresponding $y$-dependent free energy can be calculated as a sum
of node and link contributions,
\begin{equation}
  \phi(y)=\frac{1}{N} \left[ \sum_{ (i,j) \in {\cal E}}
    \phi_{i,j}^{link}(y) -\sum_{i\in{\cal V}} (d_i-1)\phi_{i}^{node}(y)
  \right]\ ,  
  \label{eq:free_energy}
\end{equation}
where $d_i$ is the degree of vertex $i$. The expressions for
$\phi_{i,j}^{link}(y)$ and $ \phi_{i}^{node}(y)$ are given explicitly
by
\begin{equation}
  \phi_{i,j}^{link}(y) = -\frac{1}{y} \ln\left( \int d^q \vec h 
  P_{i\rightarrow j}(\vec h)\ d^q\vec u Q_{j\rightarrow i}(\vec u)\ 
  \exp\left\{-y\left[ \omega(\vec h) - \omega(\vec h + \vec u )
    \right] \right\} \right)
\label{eq:link_free_energy}
\end{equation}
and by
\begin{equation}
  \phi_{i}^{node}(y) = -\frac{1}{y} \ln\left( \int \prod_{k\in V(i)}
    d^q \vec u_k Q_{k\to i}(\vec u_k) \ \exp\left\{y \omega\left(
        \sum_{k\in V(i)} \vec u_k \right) \right\} \right)\ .
  \label{eq:site_free_energy}
\end{equation}
From this free energy we can easily calculate both the complexity
$\Sigma(y)$ and the energy density $e(y)$,
\begin{equation}
\label{ene_complexity}
\Sigma(y)= -y^2 \frac{\partial \phi(y)}{\partial y} \,\,\,\,\,\,\,\, , 
\,\,\,\,\,\,\,\,
e(y)=\frac{\partial(y \phi(y))}{\partial y}\,\,\,\,, 
\end{equation}
where the complexity is defined as the logarithm of the number of
metastable states of energy $e(y)$, divided by the number $N$ of
vertices. 

Proper colorings are characterized by the limit $y\to\infty$ of the SP
Eqs.~(\ref{eq_surv_finiteyI},\ref{eq_surv_finiteyII}). In this case,
positive energy contributions are forbidden and we work directly at
zero energy. The SP equations can be brought into a much more handy
form: They are reduced to the parameters $\eta^\tau_{i \rightarrow
  j}$, i.e.~to the probabilities to have an anti-ferromagnetic message
for color $\tau$ sent from vertex $i$ to vertex $j$. Using the
vectorial notation
\begin{equation}
  \vec\eta_{i \rightarrow j} = (\eta^1_{i \rightarrow j},\eta^2_{i
  \rightarrow j}, \dots, \eta^q_{i \rightarrow j})\ ,
\end{equation}
and denoting by $V(i)\setminus j = \{k_1,k_2,\dots,k_{d_i-1}\}$ all
neighbors of $i$ different from $j$, we have the closed iteration
description 
\begin{equation}
  \vec\eta_{i \rightarrow j} =
  \vec f_{d_i-1}(\vec\eta_{k_1 \rightarrow i}, \vec\eta_{k_2
    \rightarrow 
    i},\dots,\vec\eta_{k_{d_i-1} \rightarrow i})
  \label{etaSP_compact}
\end{equation}
given component-wise by
\begin{equation}
  \eta_{i \rightarrow j}^\tau
  = \frac
  {\prod_{k \in V(i) \setminus j}(1-\eta_{k\to i}^\tau)
   -\sum_{\tau_1\neq \tau}\prod_{k \in V(i) \setminus j}
   ( 1- \eta_{k\to i}^{\tau}-\eta_{k\to i}^{\tau_1}) + \dots
   + (-1)^{q-1} \prod_{k \in V(i) \setminus j}\eta_{k\to i}^{0}}
 {\sum_{1\leq\tau_1\leq q}\prod_{k \in V(i) \setminus j}
   (1-\eta_{k\to i}^{\tau_1}) 
   - \sum_{1\leq\tau_1<\tau_2\leq q}\prod_{k \in V(i) \setminus j}
   ( 1-\eta_{k\to i}^{\tau_1}-\eta_{k\to i}^{\tau_2})+ \dots +
   (-1)^{q-1} \prod_{k\in V(i) \setminus j}\eta_{k\to i}^{0}} \ ,
\label{etaSP}
\end{equation}
for all $\tau \in \{1,\dots,q\}$. The value of $\eta ^0_{i\to j}$ can
be calculated from the normalization constraint.

The above formalism is formulated for the analysis of a single
(tree-like) graph, but it can be easily modified in order to deal with
average quantities on the random-graph ensembles ${\mathcal
  G}(N,c/(N-1))$ or ${\mathcal G}_c(N)$, see Sec.~\ref{sec:graphs}.
General considerations on the existence of a well defined
thermodynamic limit \cite{GUERRA1,FraLeo} imply the existence of a
functional probability distribution ${\mathcal Q}[Q(\vec u)]$
describing how the surveys $Q(\vec u)$ are distributed on the edges of
the graph. Noting that a $q$-component vector $\vec \eta$ is
sufficient to describe a survey $Q_{i\rightarrow j }$, we can
explicitly write ${\mathcal Q}[Q(\vec u)]$ as
\begin{equation}
 \label{eq:dist-QQ}
 {\mathcal Q}[ Q(\vec u)] = \int d^q \vec\eta \
 \rho(\vec \eta) \  
 \delta\left[ Q(\vec u) - (1-\sum_{\tau=1}^q \eta^\tau) \delta ( \vec u
   ) - \sum_{\tau=1}^q \eta^\tau
 \delta (\vec u - \vec e_{\tau}) \right]
\end{equation}
in terms of a simple $q$-dimensional probability distribution
$\rho(\vec\eta)$, with $\delta[\cdot]$ denoting a functional Dirac
distribution. The SP equations
(\ref{eq_surv_finiteyI},\ref{eq_surv_finiteyII}) have to be
interpreted in a probabilistic way: Drawing first an excess degree $d$
with probability $r_d$, cf.~Eq.~(\ref{eq:neighbors}), and then $d$
independently chosen surveys $Q_l(\vec u),\ l=1,...,d,$ from
${\mathcal Q}[Q]$, we calculate
\begin{eqnarray}
  \label{eq_surv_P}
  P_0(\vec h) &=& C_0 \int d^q
  \vec u_1\ Q_1(\vec u_1 ) \cdots  d^q \vec u_d\ Q_k(\vec u_d )\ e^{ y 
 \omega ( \sum_{l=1}^d \vec u_d )}\ \delta ( \vec h - \sum_{l=1}^d 
 \vec u_l )   
 \\
 \label{eq_surv_Q}
 Q_0(\vec u) &=& \int d^q \vec h\ P_0(\vec h )\ \delta ( \vec  u - \hat u
 ( \vec h ) )  \ .
\end{eqnarray}
The cavity equation for the functional distribution of surveys closes
by the observation that the newly generated $Q_0(\vec u)$ has to be
again a typical survey drawn from ${\mathcal Q}[Q(\vec u)]$.

Concentrating on the color-symmetric situation
$\eta:=\eta^1=...=\eta^q$, the distribution $\rho(\vec \eta)$ is
reduced to a one-dimensional $\hat\rho(\eta)$. The limit $y\rightarrow
\infty $ of the cavity equations is readily obtained:
\begin{equation}
\label{eq_sc1rsb}
\hat\rho(\eta) = 
\sum_{d=0}^{\infty} r_d
\int d\eta_1\ \hat\rho(\eta_1) \cdots d\eta_d\ \hat\rho(\eta_d)\ 
\delta (\eta - \hat f_d(\eta_1, \dots,\eta_d))
\end{equation}
with
\begin{equation}
\label{eq_self_cons_q}
\hat f_d(\eta_1,...,\eta_d) = 
\frac
{\sum_{l=0}^{q-1} (-1)^l {q-1 \choose l} \prod_{i=1}^d [1-(l+1)\eta_i]}
{\sum_{l=0}^{q-1} (-1)^l {q \choose l+1} \prod_{i=1}^d [1-(l+1)\eta_i]}
\ .
\end{equation}

It is also possible to give a closed expression for the complexity in the
COL region (notice the $p_d$ instead of $r_d$):
\begin{eqnarray}
\label{eq:sigma}
\Sigma(y =\infty) &=& 
\sum_{d=1}^{\infty}
p_d
\int d\eta_1\ \hat\rho(\eta_1) \cdots d\eta_d\ \hat\rho(\eta_d)
\ln\left(\sum_{l=0}^{q-1} (-1)^l {q \choose l+1}
\prod_{i=1}^d[1-(l+1)\eta_i] \right) \nonumber\\
&& - \frac c2 \int d\eta_1\ \hat\rho(\eta_1)\ d\eta_2
\ \hat\rho(\eta_2) \ \ln\left(1-q\eta_1\eta_2 \right).
\end{eqnarray}

\subsection{The qualitative 1RSB picture}
\label{sec:qualitative}

The formalism summarized above allows to determine not only the
location of the $q$-COL/UNCOL transition for every $q$, but the
order parameter allows also to extract important statistical
information about structure and organization of the solutions.

The most interesting point here concerns the set of solutions, seen as
a subset of all $q^N$ possible configurations $\{1,...,q\}^N$. Every
edge of the graph forbids a certain number of configurations. For
small $c$, there is, however, still an exponentially large number of
proper colorings. In addition, these are organized in a very simple
way. They are collected in one very large cluster. For any two
solutions, one can find a connecting path via other solutions, without
ever changing more than ${\cal O}(1)$ spins within one step of this
path.

\begin{figure}
\begin{center}
\hspace{-1cm}
\epsfxsize=12cm
\epsfbox{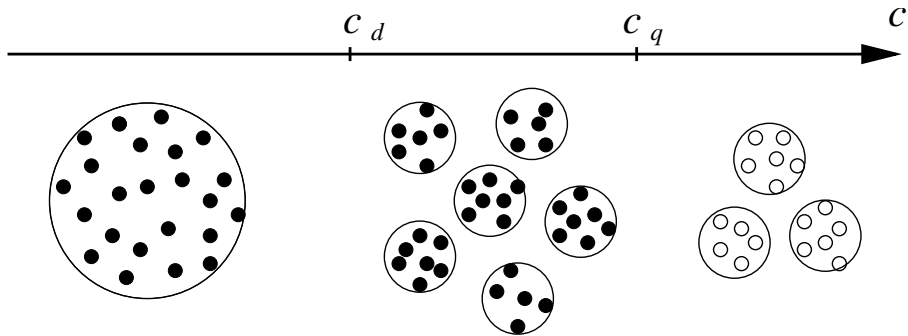}
\end{center}
\caption{A pictorial view of the solution space structure. For small
  average degree $c$, all solutions (marked by full dots) are
  collected within one large cluster. At some transition point $c_d$,
  this cluster breaks down into an exponential number of separated
  clusters. Beyond $c_q$, i.e. in the UNCOL phase, there are still
  distinct ground state clusters, but they have non-zero energy
  (marked by empty dots).}
\label{fig:clustering}
\end{figure}

This changes drastically at some average degree $c_d$: The set of
solutions is still exponentially large, but it is split into an also
exponential number of clusters. Inside each cluster, connecting paths
as described above still exist, but any two clusters have an extensive
Hamming distance from each other, cf. also Fig. \ref{fig:clustering}.
Technically the clustering threshold $c_d$ is given by the first
appearance of a non-trivial 1RSB solution. The number of clusters, or
more precisely its logarithm divided by $N$, is given by the
complexity (\ref{eq:sigma}) mentioned in the last sub-section.

Inside these clusters, the first freezing phenomena are found. A large
fraction of all vertices is frozen to one color in all solutions
belonging to the same cluster. This color changes, of course, from
cluster to cluster, since the initial model is color symmetric.
Together with this clustering of ground-states, also an exponential
number of metastable states appears. These have non-zero energy, but
they are local minima of Hamiltonian (\ref{eq_hamilton}).

The complexity of solution clusters decreases, until it vanishes at
$c_q$. Beyond this point, the graph is uncolorable. There are still
numerous ground state clusters, which have, however, non-zero energy
but zero complexity.

\section{Stability conditions of the one-step RSB solution}
\label{Sec:Stab}
There is a growing believe by now that the cavity method gives exact
results and not just approximations -- provided that the replica
symmetry is broken in the correct way. Even if a rigorous general
proof is still lacking, a number of steps forward have been made so
far in this direction. On this basis we conjecture that the results
concerning the colorability threshold, as well as many features of the
phase diagram (like for instance the existence of the clustering
phase) are {\it exact}.

In general however, the one-step RSB solution of such disordered
models has no particular reason to be correct. In some models, the
replica symmetry has to be broken infinitely many times to reach the
exact solution. In the language of the cavity method one should thus
consider an infinite hierarchy of nested clusters. This happens for
instance in the Sherrington-Kirkpatrick (SK) model~\cite{SK}, and
Talagrand has recently demonstrated rigorously that the RSB free
energy obtained in this way is exact~\cite{Tala}. On the other hand
there are models where 1RSB is not an approximation, and no further
steps of symmetry breaking are needed. This happens, e.g., in the
random-energy model~\cite{Derrida} and in the K-XORSAT
problem~\cite{XOR,XOR1} (a problem known in statistical physics as the
diluted $p$-spin model). There the 1RSB solution is known to be
rigorously exact.

Here we are going to determine whether, in the $q$-coloring problem,
one step of RSB is sufficient or whether more steps have to be taken
into account to get the correct solution. We do this by means of
analyzing the stability of the 1RSB solution against further RSB
steps.  To be more precise, we should formulate a 2RSB solution of the
model and see if the 1RSB solution is stable against a small 2RSB
perturbation \cite{note_1rsb}. This type of local stability analysis
has receive a lot of interest recently, extending the seminal work of
Elisabeth Gardner~\cite{Gardner} to finite connectivity spin glass
models~\cite{MonRicci}. It was clarified a lot in~\cite{MMM,Rivoire},
and a formalism to deal with more general finite-connectivity problems
has been established by now~\cite{Rivoire,MoRi,MerMezZec}. The
coloring problem, as considered here, will allow for nice analytical
treatments, in particular if considered on regular graphs.

\begin{figure}
\begin{center}
\hspace{-1cm}
\epsfxsize=10cm
\epsfbox{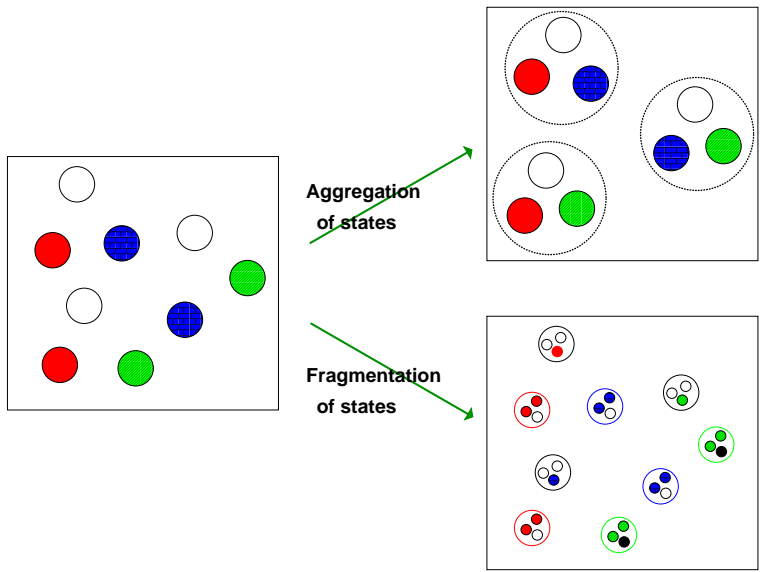}
\end{center}
\caption{A pictorial view of the two types of instabilities of the
1RSB solution, represented here for a given link $i \rightarrow j$ in
the $3$-coloring problem. In each 1RSB states (on the left), this link
carries a message of a given color, or no message at all (white
clusters here).  Going from a 1RSB to 2RSB, these states may aggregate
into bigger clusters (TYPE-I instability) or fragment in new smaller
clusters (TYPE-II instability).}
\label{InstabilityTypesFig}
\end{figure}

Let us rephrase the stability considerations for our problem,
following the notation in~\cite{MerMezZec}. Using Eq.
(\ref{eq:dist-QQ}) the cavity equations for general $y$
(\ref{eq_surv_P},\ref{eq_surv_Q}) can be rewritten in terms of the
probabilities $\eta^{\tau}, \tau=0,...,q$ only. Formally this results
in
\begin{equation}
  \label{eq:stab1}
  \eta^{\tau}_0 = C_0
  \sum_{(\sigma_1,...,\sigma_d)\to\tau} 
  \eta^{\sigma_1}_1 \cdots  \eta^{\sigma_d}_d\ \ 
  e^{y\omega(\sigma_1,...,\sigma_d)}
\end{equation}
where the sum runs over all possible combinations of input messages,
$0\leq\sigma_1,...,\sigma_d\leq q $, which induce the output message
labeled by $\tau$. Remember that the special case $y=\infty$ which
concentrates on ground states, was given explicitly in Eq.
(\ref{etaSP}).

The 1RSB solution is given by a color-symmetric distribution
$\hat\rho(\eta)$, describing $\eta$-fluctuations from link to link.
Let us, for a moment, consider an arbitrary edge $i\to j$ (together
with a direction) which is characterized by one {\it single}
$y$-dependent value of $\eta$. In the 1RSB formalism, there are many
(meta)-stable states. In each of them, this link $i\to j$ carries
exactly one warning corresponding either to one of the colors, or
being the zero message. For a randomly selected state of energy
density $e(y)$ a specific warning is found with color-independent
probability $\eta$, the trivial message appears in a fraction $1-q\eta$
of these states. This is schematically represented on the left side of
Fig.\ref{InstabilityTypesFig}.

Let us now consider two steps of RSB. There we need to take into
account the existence of clusters of states. Regarding small
perturbations of the 1RSB solution, two situations are
possible~\cite{MonRicci}: 
\begin{itemize}
\item TYPE-I: States may {\it aggregate} in configuration space,
  see Fig.\ref{InstabilityTypesFig}. In that case the order parameter
  (still for a fixed link $i \rightarrow j$) is the probability
  $g(\eta^1,...,\eta^q)$ to randomly select a cluster in which there
  is a fraction $\eta^1$ of states with message $-\vec e_1$
  transmitted from $i$ to $j$ etc. Note that it is not sufficient to
  consider a probability distribution $g(\eta)$ of a single,
  color-symmetric fraction $\eta$ since the color symmetry can be
  broken inside a cluster. This is illustrated in
  Fig.~\ref{InstabilityTypesFig}. To study the stability of the 1RSB
  solution, one should thus write the 2RSB equation, insert a
  perturbed 1RSB solution and see if it evolves back to 1RSB. In this
  case, the 1RSB solution is obtain by considering
  $g(\eta^1,...,\eta^q)=\delta(\eta^1-\eta)\cdots
  \delta(\eta^q-\eta)$. To test stability, we thus need to replace the
  Dirac peaks by narrow functions of width $\epsilon$, and to see if
  the width increases or goes to zero during the cavity iteration
  process. The analysis thus amounts to look if a small noise added to
  the solution will vanish. This is equivalent to testing convergence
  of the SP equations on a single graph.  We will refer to this kind
  of analysis as {\it noise propagation}.
\item TYPE-II: States may {\it fragment} into new states, see the
  lower right picture in Fig.\ref{InstabilityTypesFig}. In the 1RSB
  solution, a link $i\to j$ carries a message which is uniquely
  determined within one state. If now this state fragments into one
  cluster of many states, also the single message is transformed into
  a set of messages. To introduce a small perturbation of the 1RSB
  solution, we should assume that, with high probability, the states
  within one cluster are still characterized by the same message on
  link $i\to j$. In a small fraction of these states, however, also
  other messages may appear. The problem is now if or if not this
  perturbation tends to zero under iteration of the 2RSB equations.
  The stability analysis thus amounts to see if a change in one
  message (called a bug in~\cite{MerMezZec}) can propagate through the
  whole system, or if it remains localized. We will refer to this
  instability, following the terminology of~\cite{MerMezZec}, as {\it
    bug proliferation}.
\end{itemize}

\subsection{TYPE-I instability: Aggregation of states and noise
  propagation} 

Let us start with noise propagation. The stability can be computed
from the Jacobian 
\begin{equation}
  \label{first_kind}
  T^{\tau\sigma} = \frac {\partial \eta_0^\tau} {\partial
    \eta_1^\sigma} \bigg|_{\rm 1RSB}
\end{equation}
which gives the infinitesimal probability that a change in the input
probability $\eta_1^\sigma$ will change the output probability
$\eta_0^\tau$, cf. Eq. (\ref{eq:stab1}). The index 1RSB says that the
expression has to be evaluated at the 1RSB solution found within the
cavity approach. Note that we need to calculate this matrix only for
$1\leq \sigma,\tau \leq q$ since the probability of the zero-message
follows by normalization and thus does not describe an independent
quantity. After one iteration, a change of one input message of a
vertex of degree $d$ induces a change in $d-1$ outgoing messages.
The global perturbation after $n$ iterations thus concerns on average
$\(( \sum_d d r_d \))^n$ cavity messages, where $r_d$ is computed from
Eq.~(\ref{eq:neighbors}). To monitor also the strength of the
perturbation, we have to calculate
\begin{equation}
\label{eq:Tprod}
\sum_{d_1,...,d_n} d_1 r_{d_1} \cdots  d_n r_{d_n}
\ {\rm tr} \< (T_1 T_2 \cdots T_n)^2 \>_\eta
\end{equation}
where $T_1,...,T_n$ are $n$ successive $T$-matrices. The notation
$\<\cdot\>_\eta$ denotes the average over the external messages to the
nodes $1,...,n$. To be more precise we should write $T_i =
T_{d_i}(\eta_{i,1},...,\eta_{i,d_i})$, i.e. $T_i$ depends on $d_i$
incoming probabilities $\eta_{i,1},...,\eta_{i,d_i}$, with $d_i$ being
the excess degree of vertex $i$ distributed according to $r_{d_i}$.
For $i=1$, all messages are external. They have to be generated
independently from the 1RSB order parameter $\hat\rho(\eta)$.  For
$i>1$, the first message described by $\eta_{i,1}$ results from node
$i-1$ according to Eq.~(\ref{eq:stab1}). The other inputs
$\eta_{i,2},...,\eta_{i,d_i}$ are again external and thus
independently distributed with $\hat\rho(\eta)$. Note that the
$\eta$-averaged matrix $\< (T_1 T_2 ...  T_n)^2 \>_\eta$ still depends
on the random excess degrees $d_1,...,d_n$. This has to be taken into
account in the sum in (\ref{eq:Tprod}).

A simple way to calculate the trace for $n\gg1$ is to consider the
biggest eigenvalue of the matrix $\< (T_1 T_2 ...  T_n)^2 \>_\eta$,
which allows us to use a single number to follow the perturbation.
This, however, becomes very simple for $q$-COL. Let us rewrite the
Jacobian:
\begin{eqnarray}
  \label{eq:T-matrix}
  T = \left[
    \begin{array}{ccccccc} 
      \frac {\partial \eta_0^1} {\partial \eta_1^1} &&
      \frac {\partial \eta_0^2} {\partial \eta_1^1} && ... &&
      \frac {\partial \eta_0^q} {\partial \eta_1^1} \\ \\
      \frac {\partial \eta_0^1} {\partial \eta_1^2} &&
      \frac {\partial \eta_0^2} {\partial \eta_1^2} && ... &&
      \frac {\partial \eta_0^q} {\partial \eta_1^2} \\ \\
      ... && ... && ... && ... \\ \\
      \frac {\partial \eta_0^1} {\partial \eta_1^q} &&
      \frac {\partial \eta_0^2} {\partial \eta_1^q} && ... &&
      \frac {\partial \eta_0^q} {\partial \eta_1^q} 
    \end{array}
  \right]_{\rm 1RSB} \ .
\end{eqnarray}
Evaluated at the color-symmetric 1RSB solution, this matrix has only
two different entries: All diagonal elements are equal, and all
non-diagonal elements are equal. As an immediate consequence all
Jacobians commute and are thus simultaneously diagonalizable. The
matrix $T$ has only two distinct eigenvalues,
\begin{equation}
  \label{jac_eig}
  \begin{cases}
    \lambda_1 = \((  \frac {\partial \eta_0^1} {\partial \eta_1^1} -
    \frac {\partial \eta_0^1} {\partial \eta_1^2} \)) \Big|_{1RSB}
    \\  \\
    \lambda_2 =\(( \frac {\partial \eta_0^1} {\partial \eta_1^1} +
    (q-1)\ \frac {\partial \eta_0^1} {\partial \eta_1^2} \))
    \Big|_{1RSB}  
\end{cases}
\end{equation}
The second eigenvalue belongs to the homogeneous eigenvector
$(1,1,...,1)$. It describes a fluctuation changing all $\eta_1^\tau,
\tau=1,...,q,$ by the same amount, i.e. a fluctuation maintaining the
color symmetry. The first eigenvalue is $(q-1)$-fold degenerate,
eigenvectors are spanned by $(1,-1,0,...,0),\ (0,1,-1,0,...,0),\ ...,\
(0,...,0,1,-1)$. The corresponding fluctuations explicitly break the
color symmetry, and they are in fact found to be the critical ones for
the instability of the 1RSB solution.

For zero-energy ground states in the COL phase, where $y=\infty$, one
may proceed using the closed analytical expression~(\ref{etaSP}) for
the iteration.  It is possible to explicitly write down both
eigenvalues. We will see in section~\ref{Sec:High-q} how to derive
asymptotic results at large $q$. Finally, to monitor the perturbation,
one should consider the dependence of
\begin{equation}
  \label{stabilita2}
  \lambda_{TYPE-I}(n) =  
  \sum_{d_1,...,d_n} d_1 r_{d_1} \cdots  d_n r_{d_n}
  \ \lambda_{\< (T_1 \cdots T_n)^2 \>_\eta} 
\end{equation}
on the number $n$ of iteration steps. There $\lambda_{\< (T_1 T_2 ...
  T_n)^2 \>_\eta}$ denotes the biggest eigenvalue of matrix $\< (T_1
T_2 ... T_n)^2 \>_\eta$. It can be computed as a product of
eigenvalues being either all of type $\lambda_1$ or all of type
$\lambda_2$. The 1RSB solution is stable against TYPE-I perturbations
if and only if $\lambda_{TYPE-I}(n)$ decays to zero in the large-$n$
limit.

\subsection{TYPE-II instability: Fragmentation of states and bug
  proliferation} 

As discussed above, we have to consider bug proliferation in order to
study the second type of instability, i.e.~the instability with
respect to fragmentation of states into clusters of states. Suppose
that a message of type $\sigma$ is turned into another message
$\tilde\sigma$. This is called a bug in \cite{MerMezZec}. We
suppose it to happen with small probability
$\pi^{\sigma\to\tilde\sigma} \ll 1$. As a consequence, some output
messages may change, and the bug propagates. The system is unstable
with respect to TYPE-II perturbations if such a bug propagates through
the whole system.

Considering small perturbations, we can work in the linear response
regime. For a generic change of the first input message in
Eq.~(\ref{eq:stab1}), we may calculate the probability of changing the
output message:
\begin{equation}
  \label{eq:type2}
  \pi^{\tau\to\tilde\tau}_0 = C_0 
  \sum_{\substack { (\sigma,\sigma_2,...,\sigma_d) \to \tau \\
   (\tilde\sigma,\sigma_2,...,\sigma_d) \to \tilde\tau } } 
  \pi^{\sigma\to\tilde\sigma}_1 
  \eta^{\sigma_2}_2 \cdots  \eta^{\sigma_d}_d \ \ 
  e^{y\omega(\tilde\sigma,\sigma_2,...,\sigma_d)} 
\end{equation}
This defines a matrix $V$ with entries
\begin{equation}
\label{eq:V}
  V_{\tau\to\tilde\tau,\sigma\to\tilde\sigma} = \frac
  {\partial\pi^{\tau\to\tilde\tau}_0 }{\partial
    \pi^{\sigma\to\tilde\sigma}_1 } \ ,
\end{equation}
evaluated at the 1RSB solution. Since a message may have $q+1$
different states and we are considering actual changes in messages,
$V$ is a square matrix of dimension $q(q+1)$. We study it in
Appendix~\ref{appendix:TypeII} and show that its biggest eigenvalues
can be written, at any given $y$, as a function of $q$ and
$\eta_1,...,\eta_d$ as
\begin{equation}
  \lambda = (q-1)\ {\cal{A}}(q,\{{\eta\}}) + \sqrt{{\cal{B}}(q,\{{\eta\}})
    \ {\cal{B'}}(q, \{{\eta\}})}
\end{equation}
where ${\cal A,\ B}$ and ${\cal B}'$ are explicitly given in
Eq.~(\ref{expression_abc}) in App.~\ref{appendix:TypeII}. In the
zero-energy case, i.e.~for $y \rightarrow \infty$, this expression
simplifies to
\begin{equation}
\lambda = (q-1) {\cal{A}}_{y=\infty}(q,\{{\eta\}})
\end{equation}
where ${\cal{A}}_{y=\infty}(q,\{{\eta\}})$ is given by
\begin{equation}
\label{a_of_q}
{\cal{A}}_{y=\infty}(q,\{{\eta\}}) = \frac
{\sum_{l=0}^{q-2} (-1)^l {q-2 \choose l} \prod_{i=2}^d [1-(l+2)\eta_i] }
{\sum_{l=0}^{q-1} (-1)^l {q \choose l+1} \prod_{i=1}^d
[1-(l+1)\eta_i] }\ .
\end{equation}

The instability with respect to bug proliferation has to be determined
from a product of $n$ such matrices. Proliferation is monitored, using
again Eq.~(\ref{eq:neighbors}), by
\begin{equation}
  \sum_{d_1,...,d_n} d_1 r_{d_1} \cdots  d_n r_{d_n}
 \ {\rm tr} \< V_1 V_2 \cdots V_n \>_\eta 
\end{equation}
or, using the eigenvalue notation, by
\begin{equation}
  \label{fluctu_stab_II}
  \lambda_{TYPE-II}(n) =   
  \sum_{d_1,...,d_n} d_1 r_{d_1} \cdots  d_n r_{d_n}
 \ \lambda_{\< V_1 V_2 \cdots V_n \>_\eta} 
\end{equation}
where $\lambda_{\< V_1 V_2 ... V_n \>_\eta}$ is the biggest eigenvalue
of matrix $\< V_1 V_2 ... V_n \>_\eta$. The averages are performed in
complete analogy to averages in the TYPE-I case. The 1RSB solution is
stable if and only if this eigenvalue decays to zero for $n\to\infty$.

\section{Study of the TYPE-I instability: Noise propagation}
\label{Sec:TypeI}

Using the criteria derived in the previous section, we can now study
the stability of the 1RSB solution for the ground states of the
coloring problem, i.e. we concentrate first on $y\to\infty$. We will,
in the following, call $c_{SP}$ the connectivity beyond which 1RSB is
TYPE-I unstable because, as we will see, the SP equations do not
converge anymore on a single graph for $c>c_{SP}$.

\subsection{Regular random graphs}

In the case of fixed-connectivity graphs all sites are equivalent, so
$\eta$ does not fluctuate from edge to edge. Therefore, the recursion
equation (\ref{eq_self_cons_q}) can be simplified to
\begin{equation}
  \label{eq_self_cons_q_factorized}
  \eta = \hat f_k (\eta) = 
  \frac{\sum_{l=0}^{q-1} (-1)^l {q-1 \choose l} [1-(l+1)\eta]^k}
  {\sum_{l=0}^{q-1} (-1)^l {q \choose l+1} [1-(l+1)\eta]^k }
\end{equation}
with $k=c-1$. This equation can be easily solved using basic numerical
tools. Similar simplifications arise in the computation of the
complexity $\Sigma$. Within the 1RSB formalism, it is therefore very
easy to derive $\eta(q,c)$, to compute the corresponding complexity,
and thus to determine if the graph is in the COL phase or not. One
finds that that at small connectivities the solution is trivial,
i.e.~$\eta=0$. For $c \geq c_d$ the clustering RSB phenomenon occurs,
and for $c \geq c_{q}$ the graph becomes uncolorable. To test the
validity of the 1RSB ansatz we are now going to apply the criteria
derived previously. Due to the fixed vertex degree, there is no need
to average over the site distribution. The stability criterion thus
simplifies considerably to
\begin{equation}
  \label{Crit:Fac_Type_I}
  \lambda_{TYPE-I}(1) = 
  k  \((  \frac {\partial \eta_0^1} {\partial \eta_1^1} - \frac
  {\partial \eta_0^1} {\partial \eta_1^2} \))^2 \Big|_{1RSB} <1 \ . 
\end{equation}
In Tab.~\ref{mine}, we summarize the values of $\eta$ and $\Sigma$, as
well as stability criteria, for $q=3,\ 4$ and $5$. The instability of
TYPE-I appears at high connectivity, in the UNCOL phase. So it turns
out to be irrelevant in the COL phase. This turns out to be true for
arbitrary number $q$ of colors, see Sec.~\ref{Sec:High-q} for the
asymptotic case. However, this instability is directly relevant for
the behavior of the SP algorithm on single graphs: When $c \geq c_{SP}
$, SP stops to converge on a single graph. This is actually what the
Jacobian (\ref{first_kind}) implies first of all. We have verified
this numerically for 3-COL, SP does not converge on regular graphs of
degree equal to or larger than 6.
\begin{table}[htbp]
\begin{tabular}{|c||c|c|c|c|}
\hline
$q,~c$ & phase & $\eta$ & $\Sigma$ & $\lambda_{TYPE-I}$ \\
\hline
\hline
$q=3,\ c=4$ & COL  & & & \\
$q=3,\ c=5$ & COL~1RSB  & 0.279 & 0.037 & 0.688 \\
$q=3,\ c=6$ & UNCOL~Instable & 0.318 & -0.137 & 1.103 \\
$q=3,\ c=7$ & UNCOL~Instable & 0.327 & -0.329  & 1.420\\
\hline
$q=4,\ c=8$ & COL & & & \\
$q=4,\ c=9$ & COL~1RSB & 0.223 & 0.040 & 0.764 \\
$q=4,\ c=10$ & UNCOL~1RSB & 0.236 & 0.03 & 0.86 \\
$q=4,\ c=11$ & UNCOL~Instable & 0.242 & -0.216 & 1.02 \\
\hline
$q=5,\ c=12$ & COL & & & \\
$q=5,\ c=13$ & COL~1RSB & 0.1759 & 0.0960 & 0.5605 \\
$q=5,\ c=14$ & COL~1RSB & 0.1853 & 0.00380 & 0.6788 \\
$q=5,\ c=15$ & UNCOL~1RSB & 0.1902 & -0.0952 & 0.775 \\
$q=5,\ c=16$ & UNCOL~1RSB & 0.1932 & -0.1982 & 0.8622 \\
$q=5,\ c=17$ & UNCOL~1RSB & 0.1952 & -0.3037 & 0.942 \\
$q=5,\ c=18$ & UNCOL~Instable & 0.1966 &  -0.4109 & 1.017 \\
\hline
\end{tabular}
\caption{\label{mine}
  Computation of $\eta$, $\Sigma$ and $\lambda_{TYPE-I}$ for $q=3,4$
  and $q=5$ and various values of the connectivity on regular random
  graphs}  
\end{table}

\subsection{Fluctuating-connectivity random graphs: Erd\"os-R\'enyi
  ensemble} 

\begin{figure}
\begin{center}
\includegraphics[width=8cm]{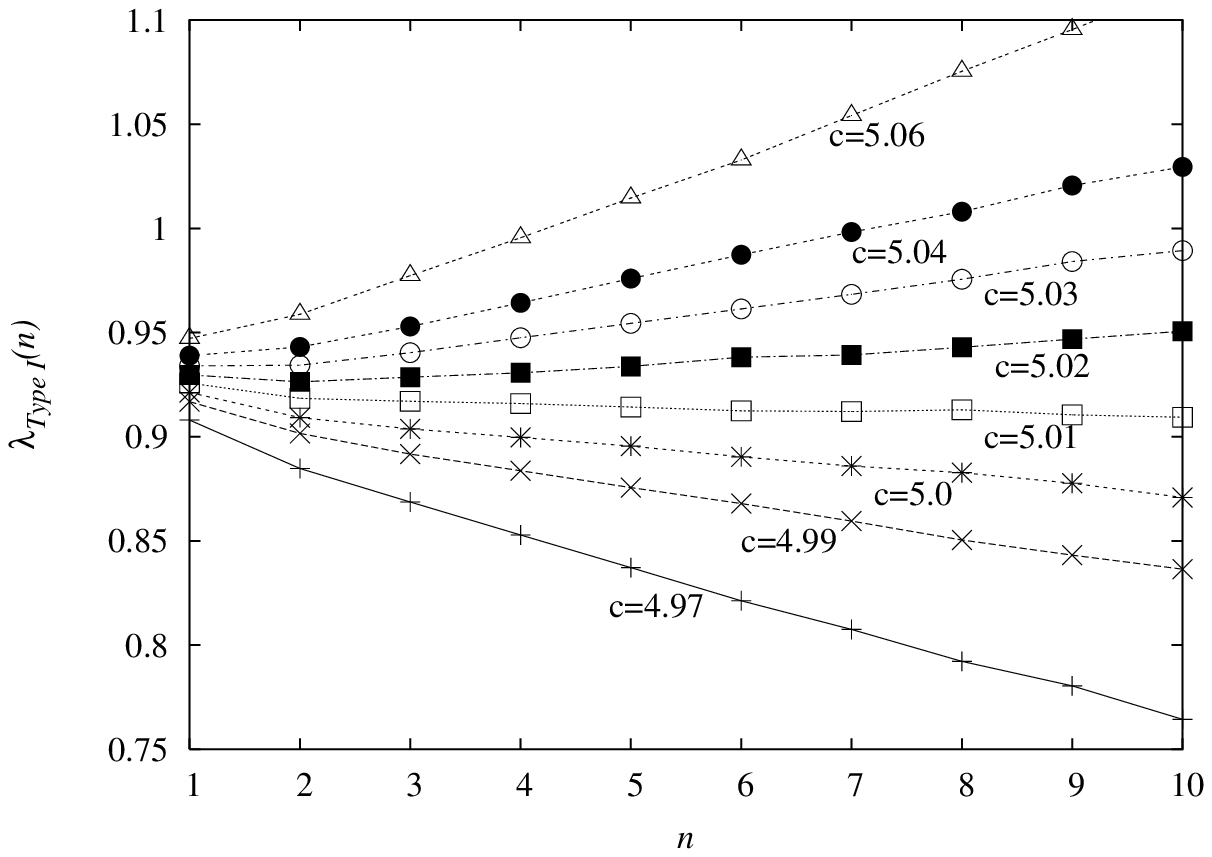} 
\includegraphics[width=8cm]{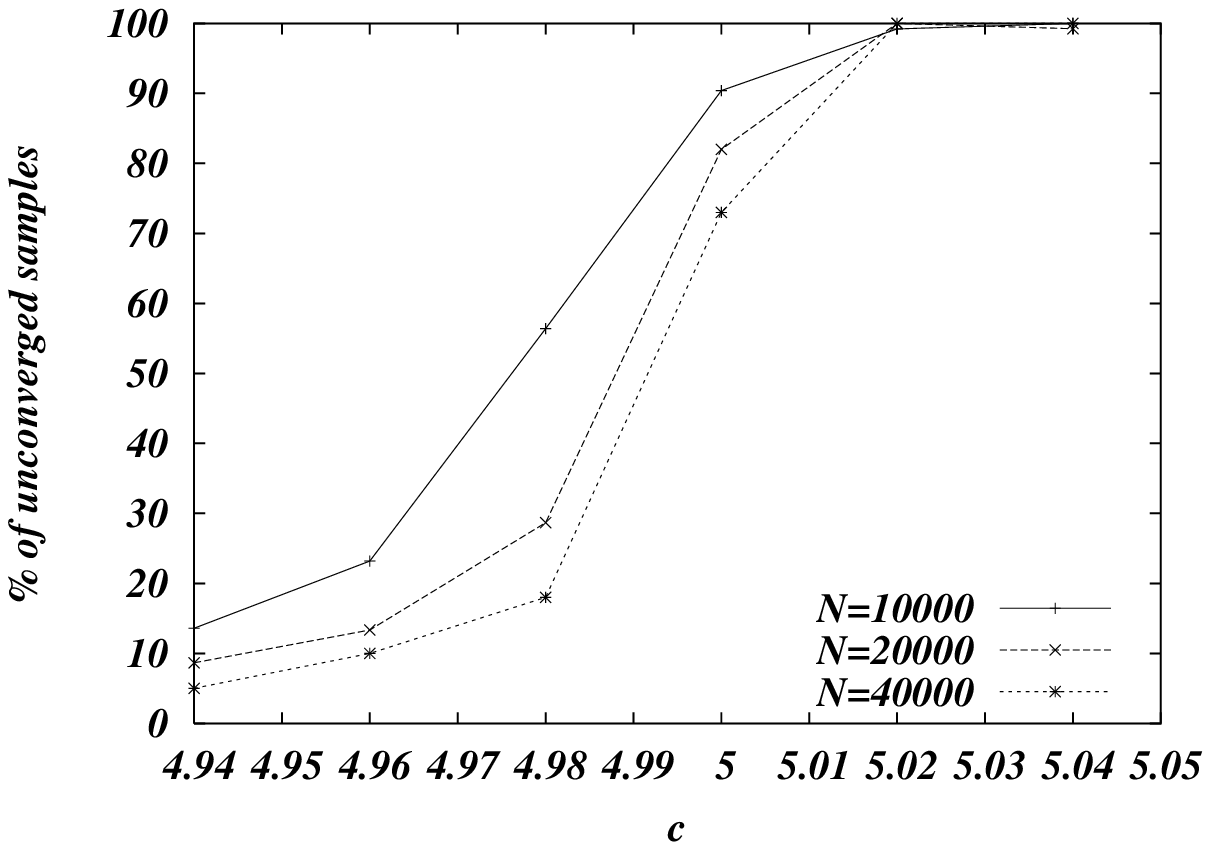}
\end{center}
\caption{Analytic and numerical computation of the location of the
TYPE-I instability point in the $3$-coloring on fluctuating
connectivities graphs. We find $c(3)_{SP} \simeq  5.01$ in both
cases, which confirms well that SP equation stop to converge when
$c>c_{SP}$.}
\label{InstabilityTypesI_q3}
\end{figure}

Let us now turn to Erd\"os-R\'enyi random graphs. Sites are no more
equivalent and the order parameters $\eta$ fluctuate from link
to link. We have to be very careful in the disorder average in the
stability criterion, a detailed description was given above. The
TYPE-I instability amounts to see if a small change in $\eta$
propagates through the whole system. We saw in section~\ref{Sec:Stab}
that this can be computed by the study of a Jacobian that describes
the propagation of a perturbation after one iteration. The global
perturbation after $n$ iteration is monitored by the sum of the
squares of the perturbed cavity-bias, which behaves like
\begin{equation} 
  \lambda_{TYPE-I}(n) = 
  \sum_{d_1,...,d_n} d_1 r_{d_1} \cdots  d_n r_{d_n}
 \ \lambda_{\< (T_1 T_2 \cdots
    T_n)^2 \>_\eta}  
\end{equation} 
where $\lambda_{\< (T_1 \cdots T_n)^2 \>_\eta}$ is the maximal
eigenvalue of the $\eta$-average of matrix $ (T_1 \cdots T_n)^2 $. The
system is stable if $\lambda_{TYPE-I}(n)$ goes to zero with $n$, and
instable if it diverges. Let us concentrate for a moment on 3-COL:
Using the SP equation on single graphs, we saw that the iteration is
not converging anymore for $c>5.01$. Evaluating the stability
criterion, we were able to reproduce this number analytically (see
Fig.\ref{InstabilityTypesI_q3}). Both methods agrees exactly.
We have summarized our result for $c_{SP}$ in the last column of
Tab.~\ref{tab:results_complete}.  Again, the conclusion is that the
TYPE-I instability is irrelevant for the thermodynamics of the
system.

\section{Study of the TYPE-II instability: Bug proliferation}
\label{Sec:TypeII}

We turn now to the instability of TYPE-II. Like before, we start with
regular random graphs, and then we move to Erd\"os-R\'enyi graphs.
In~\cite{MonRicci}, it is argued that, if 1RSB states are instable,
then these states should be described using a full RSB ansatz, and
therefore are marginally stable~\cite{Cyrano}.  The connectivity at
which the instability appears was therefore called $c_m$ for
$c_{marginal}$. Although it is not yet clear if full RSB is needed
in the case 1RSB is unstable, we follow this notation and call
$c_m$ the connectivity at which the TYPE-II instability shows up.

\subsection{Regular random graphs}

Using again the homogeneous solution, it is very easy to write the
stability criterion. It reads (see appendix~\ref{appendix:TypeII_A})
\begin{equation}
\label{Crit:Fac_II}
\lambda_{TYPE-II} = k(q-1) \frac {\sum_{l=0}^{q-2} (-1)^l {q-2
    \choose l} [1-(l+2)\eta]^{k-1} } {\sum_{l=0}^{q-1} (-1)^l {q
    \choose l+1} [1-(l+1)\eta]^k } <1 
\end{equation}
with $k=c-1$. Applying this criterion, it turns out that the
instability does not appear at low $q$. In fact, for $q<6$, the
ground state in the COL region is always either RS or 1RSB, no
instability toward 2RSB is observed. However, when $q=6$, the lowest
connectivity for which RSB appears turns out to be instable. More
generally, there is an unstable zone growing with $q$ before the
stable 1RSB region is reached. We summarize our result for
small $q$ in Tab.~\ref{tab:results_fixed}. Note that $c_d$ is
calculated in 1RSB approximation, if the latter is instable its
position may change due to more-step RSB.
\begin{figure}
\begin{center}
\includegraphics[width=8cm]{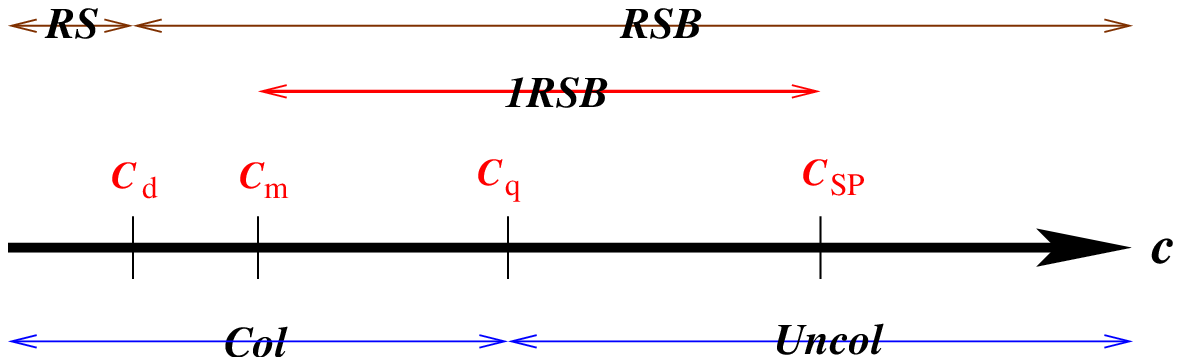} 
\caption{Schematic phase diagram for the coloring transition at $e=0$.
  At small connectivity, when $c<c_{d}$ the problem is COL and the
  space phase is trivial. Then, for $c \geq c_{d}$ the phase space is
  broken into many states and RSB appears. In this RSB phase there is
  a one-step RSB zone $c_{m}\leq c<c_{SP}$ in which detailed
  computations are possible. The COL/UNCOL transition appearing at
  $c_{q}$ is always found in this zone. Note that for some
  connectivities, $c_{d}$ and $c_{m}$ are equal.}
\label{phase_shema}
\end{center}
\end{figure}
\begin{table}[htbp]
\begin{tabular}{|c||c|c|c|c|}\hline
q & $c_{d} $ & $c_{m} $ & $c_{q}$ & $c_{SP}$ \\
\hline \hline
3 & 5  & 5 & 6 &  6  \\  
\hline
4 & 9 & 9  & 10 &  11  \\  
\hline
5 & 13& 13 & 15 &  18 \\  
\hline
6 & 17  & 18 & 20 & 27 \\  
\hline
7 & 21 & 22  & 25 & 38 \\  
\hline
8 & 26 & 27 & 31 & 51 \\  
\hline
9 & 31  & 32 & 37 & 66  \\  
\hline
10 & 36  & 37 & 44 & 83  \\  
\hline
11 & 41 & 43 & 50 & 102 \\  
\hline
12 & 46 & 48 & 57 & 123 \\  
\hline
\end{tabular}
\caption{Critical connectivities of the coloring problem on fixed
connectivities random graphs.}
\label{tab:results_fixed}
\end{table}

\begin{figure}
\begin{center}
\includegraphics[width=10cm]{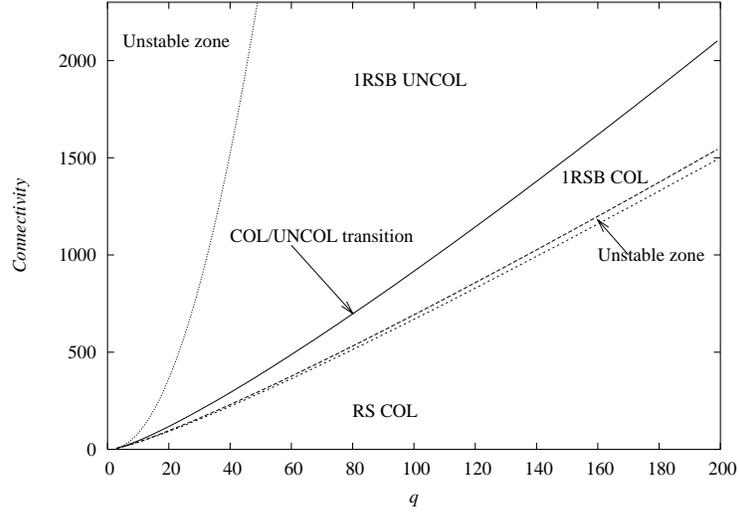} 
\end{center}
\caption{Phase diagram of the q-coloring problem on random graphs with
  fixed connectivities, for 3 to 200 colors. From bottom to top we
  have four line corresponding to $c_d$, $c_m$,$c_q$ and $c_{SP}$. The
  critical line $c_q$ that separates the COL and the UNCOL region is always in
  the stable zone, where we conjecture our results to be exact. Note
  however the small, but slowly growing, zone in the COL region where
  the 1RSB computation is unstable.}
\label{PD_fixed}
\end{figure}

The one-step solution is thus {\it not} always stable in COL phase. It
seems, however, to be correct close to the COL/UNCOL threshold, the
result for the latter is therefore conjectured to be exact. The global
situation is schematically represented in the Fig.~\ref{phase_shema}.
We have derived the solution of this problem up to $q=200$, which
allows us to draw a quite complete phase diagram for the problem on
finite connectivity random graphs, see Fig.~\ref{PD_fixed}.
From this phase diagram, two important points have to be noticed: 
\begin{itemize}
\item The critical line for the COL/UNCOL transition is always in the
  1RSB stable zone, and therefore we believe it to be be an exact
  result.
\item There are two zones where the 1RSB solution is unstable. This is
  happening at high connectivity in the UNCOL phase, due to a TYPE-I
  instability, and also in a small, though growing (with the number of
  colors) zone between the RS COL and the 1RSB COL phases, due to a
  TYPE-II instability.
\end{itemize}

\subsection{Erd\"os-R\'enyi random graphs}
\label{sec:II_ER}

\begin{figure}
\begin{center}
\includegraphics[width=10cm]{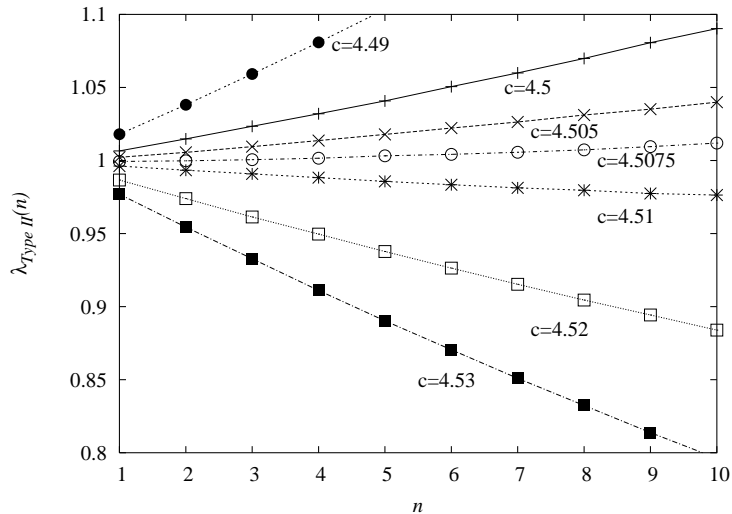} 
\end{center}
\caption{Computation of the location of the TYPE-II 
  instability point in 3-COL on Erd\"os-R\'enyi graphs.  One finds
  $c_{m} \simeq  4.51$.}
\label{InstabilityTypesII_q3}
\end{figure}

Let us now go back to the Erd\"os-R\'enyi ensemble. As mentioned
above, the instability is monitored by
\begin{equation}
  \label{Crit:Fac_II_bis}
  \lambda_{TYPE-II}(n) = (q-1)^n   
 \sum_{d_1,...,d_n} d_1 r_{d_1} \cdots  d_n r_{d_n}
 \left\< 
    \prod_{i=1}^n 
    \frac {\sum_{l=0}^{q-2}
      (-1)^l {q-2 \choose l} \prod_{j=2}^{d_i} [1-(l+2)\eta_{i,j}] }
    {\sum_{l=0}^{q-1} (-1)^l {q \choose l+1} \prod_{j=1}^{d_i}
      [1-(l+1)\eta_{i,j}] 
    } 
  \right\> <1 
\end{equation}
where the $d_i$ has the characteristic Poissonian distribution
$r_{d_i}$ of average $c$. As already explained above, the
$\eta$-values are given in the following way: For the first node
$i=1$, all values of $\eta_{1,1},...,\eta_{1,d_1}$ are drawn
independently from $\hat\rho(\eta)$. This allows to calculate the
first factor in the above product, but also the first input message
$\eta_{2,1} = \hat f_{d_1} (\eta_{1,1},...,\eta_{1,d_1})$ to the
second node. The other $d_2-1$ input messages
$\eta_{2,2},...,\eta_{2,d_2}$ are drawn again from $\hat\rho(\eta)$.
This is repeated for the other vertices: The input message on the link
coming from the previous node is induced via the SP equation, the
others are drawn randomly.  Using again the population dynamic
solution of the coloring problem, it is easy to perform this average.
For 3-COL, e.g., we find the instability to be present for
connectivities $c<c_{m} \simeq 4.51$, see
Fig.~\ref{InstabilityTypesII_q3})
We have repeated this study for $q=4$ and $5$. Our results are
summarized in Tab.~\ref{tab:results_complete}. As opposed to what
happens in the fixed-connectivity case, there is an unstable 
interval close to the clustering transition at $q=3,4$.
\begin{table}[htbp]
\begin{tabular}{|c||c|c|c|c|}\hline
q & $c_{d} $ & $c_{m} $ & $c_{q}$ & $c_{SP}$ \\
\hline \hline
3 & 4.42  & 4.51 & 4.69 &  5.01  \\  
\hline
4 & 8.27  & 8.34  & 8.90 &  10.21  \\  
\hline
5 & 12.67 & 12.67 & 13.69 & 17.1 \\  
\hline
\end{tabular}
\caption{Critical connectivities of the coloring problem on
Erd\"os-R\'enyi graphs.}
\label{tab:results_complete}
\end{table}
For $q=5$, however, it is always stable. As we will see in the next
section, at larger connectivities the unstable zone re-appears.
Again, the most important conclusion is that the critical values for
the COL/UNCOL transition are in the stable zone. We thus conjecture
that the connectivities for the COL/UNCOL transition are {\it exact}.

\section{Asymptotic threshold values in the high-$q$ limit}
\label{Sec:High-q}

Whereas the precise values for transitions and stability regions of
the 1RSB approach have to be determined by numerically solving the
cavity equations, and by plugging this solution into the stability
criteria, an asymptotic analysis can be carried through analytically
in the limit of a large number of colors, i.e. for $q\gg 1$.

As an astonishing result, Poissonian and regular random graphs have
the same asymptotic behavior, at least what concerns the dominant
contributions. Degree fluctuations play a role only at low-order
contributions which, due to the complexity of their determination, 
are not considered here. To lighten the presentation, we consider
therefore only the case of regular graphs, i.e.~of equal vertex
degrees $c=k+1$ for all vertices.

As discussed above, in this case the 1RSB solution is factorized and
thus completely determined by a single edge- and color-independent
$\eta$. A basic numerical observation in this context is that, for
$q\gg 1$, this $\eta$ approaches $1/q$, whereas the probability
$\eta^0=1-q\eta$ of sending a trivial message vanishes asymptotically.
We can therefore write $q \eta= 1-g(q,c)$, where the positive, but
still unknown function $g(q,c)$ will be shown to vanish in the
large-$q$ limit.

Note that, in the case of fluctuating vertex degrees, also $\eta$ will
fluctuate according to some non-trivial $\rho(\eta)$. For large $q$,
this distribution concentrates, however, more and more sharply around
its average value. A careful analysis of the influence of these
fluctuations shows that they contribute only to sub-dominant terms in
the threshold values, i.e. they can be neglected to the considered
orders in $q$.

\subsection{The clustering transition in 1RSB approximation}

The clustering transition is characterized by the onset of a
non-trivial solution of the survey propagation equation
(\ref{eq_sc1rsb}). To determine this point, we consider a graph of
constant degree $c=k+1$ with
\begin{equation}
  \label{eq:kd_asymp}
  k = q [ \ln q + \ln \ln q + \alpha]\ .
\end{equation}
There $\alpha$ is an arbitrary constant, i.e. in the large-$q$ limit
$k$ is determined down to ${\cal O}(q)$. Doing so, we zoom directly
into the region where the dynamical transition appears: We look for an
$\alpha_d$ at which the first non-trivial $\eta$ appears as a solution
of
\begin{equation}
\label{eq:eta_asymp}
\eta = 
\frac{\sum_{l=0}^{q-1} (-1)^l {q-1 \choose l} [1-(l+1)\eta]^k}
{\sum_{l=0}^{q-1} (-1)^l {q \choose l+1} [1-(l+1)\eta]^k }\ .
\end{equation}
As a first step we realize that, for $\eta\simeq 1/q$ and $k$ as given
in (\ref{eq:kd_asymp}), the sums in both the numerator and the
denominator are dominated by values of $l\ll q$, where we can replace
\begin{eqnarray}
  \label{eq:replace}
  [1-l\eta]^k &\simeq& \exp\{ -\eta k l\}
  \nonumber\\
  &\simeq& \exp\{ -[1-g(q,\alpha)]\ [ \ln q + \ln \ln q + \alpha]\ l \}
  \nonumber\\
  &\simeq& \left[ \frac 1{q\ln q}\ q^{g(q,\alpha)}\ e^{-\alpha} 
  \right]^l 
\end{eqnarray}
up to neglectable corrections. Note the change in notation from
$g(q,c)$ to $g(q,\alpha)$. Plugging this into Eq.
(\ref{eq:eta_asymp}), we find
\begin{eqnarray}
  \label{eq:gqc}
  1-g(q,\alpha) &\simeq& 
  \frac 1{\ln q}\ q^{g(q,\alpha)}\ e^{-\alpha} \
  \frac{\exp\left\{- \frac 1{\ln q}\ q^{g(q,\alpha)}\ e^{-\alpha} \right\}}
  {1-\exp\left\{- \frac 1{\ln q}\ q^{g(q,\alpha)}\ e^{-\alpha} \right\}}
  \nonumber\\
  &\simeq& 1 - \frac 1{2 \ln q}\ q^{g(q,\alpha)}\ e^{-\alpha} \ .
\end{eqnarray}
This equation closes for $g(q,\alpha) = \gamma(\alpha) / \ln q$, with
the condition
\begin{equation}
  \label{eq:gamma}
  \gamma(\alpha) e^{-\gamma(\alpha)} = \frac 12 e^{-\alpha}\ .
\end{equation}
The maximum of the left-hand side is $1/e$ for $\gamma=1$, i.e. a real
solution for $\gamma(\alpha)$ exists if and only if $\alpha>1-\ln 2$.
We thus find the dynamical transition in the 1RSB approach at
\begin{eqnarray}
  \label{eq:kd_eta_asymp}
  c_d &=& q\ \left[ \ln q + \ln \ln q +1 -\ln 2 + o(1)\right]\\
  \eta_d &=& \frac 1q \left[ 1- \frac 1{\ln q} + o(\ln q ^{-1})
  \right] \ . \nonumber
\end{eqnarray}
The result is equally valid for Poissonian random graphs. This can be
understood immediately: Degree fluctuations are of ${\cal O}(\sqrt c)=
{\cal O}(\sqrt{ q\ln q})$, and they are thus neglectable compared to
the contributions in Eq.~(\ref{eq:kd_asymp}).

\subsection{The COL/UNCOL transition in 1RSB approximation}

From the upper bound of {\L}uczak and the lower one of Achlioptas and
Naor we know that
\begin{equation}
  \label{eq:cd_asymp}
  c_q \simeq 2 q \ln q - \ln q + {\cal O}(1)\ .
\end{equation}
We are going to re-derive this result from the cavity equations, and we
also determine the previously unknown ${\cal O}(1)$-contribution. 
Observing that, under this scaling of the degree, the contributions in 
the self-consistency equation (\ref{eq:eta_asymp}) behave as
$[1-l\eta]^k \sim q^{-2l}$, we find
\begin{eqnarray}
  \label{eq:etac_asympt}
  \eta &=& \frac 1q \left( 1 - \frac 12 \ (q-1) \ \left[ 
  \frac{1-2\eta}{1-\eta} \right]^k  \right) + {\cal O}(q^{-3})
  \nonumber\\
  &=& \frac 1q \left( 1 - \frac 12 \ (q-1) \ \exp\left\{ 
  -k\eta - \frac 32 k \eta^2\right\}  \right) + {\cal O}(q^{-3})\ .
\end{eqnarray}
This equation is solved by
\begin{equation}
  \label{eq:etac_sol}
  \eta = \frac 1q - \frac 1{2q^2} + \frac {3 \ln q}{2 q^3}
  + {\cal O}(q^{-3})\ ,
\end{equation}
independently on the ${\cal O}(1)$-term in (\ref{eq:cd_asymp}).
This value has to be plugged into the expression for the complexity,
\begin{equation}
  \label{eq:compl_asympt}
  \Sigma = \ln\left( \sum_{l=0}^{q-1} (-1)^l {q \choose l+1}
    [1-(l+1)\eta]^c \right) - \frac c2 \ln(1-q\eta^2) \ ,
\end{equation}
in order to determine the critical point $c_q$ of the COL/UNCOL
transition from the vanishing of $\Sigma$. Keeping only the three
dominant terms in $q$, we immediately find
\begin{equation}
  \label{eq:cd_result}
  c_q \simeq 2 q \ln q - \ln q - 1 + o(1)\ .
\end{equation}
The result is equally valid for Poissonian graphs, and it coincides
precisely with the upper bound of {\L}uczak, i.e. his improved
annealed approximation is asymptotically exact. It is, however, also
only one larger than the conjectured lower bound by Achlioptas and
Naor. In this way we see that the 1RSB approach is not only consistent
with the mathematical bound, but allows for a more precise (even if
not rigorous) determination of the threshold value.

\subsection{Asymptotic stability}

To check the validity of the above results, we have to certify that
the 1RSB solution is locally stable. Again we discuss only the case of
the regular random graph of degree $c=k+1$, but the results do not
differ in the Poissonian case.

\subsubsection{Instability of TYPE-I}

Let us first start with the instability of TYPE-I. To do so, we should
compute the eigenvalues of the stability matrix derived in
Sec~\ref{Sec:TypeI}.  For average connectivities $c \gg q$, the
recursion equation (\ref{etaSP}) is dominated by the first
contribution in both the numerator and the denominator, leading to
$\rho(\eta)=\delta(\eta-1/q)$ in leading order. Corrections are
exponentially small in $c/q$ and can thus be neglected.  More
precisely, taking the asymptotic $c \gg q$ results of
Eq.~(\ref{etaSP_compact}) and then using the fact that all colors on
all branches except the perturbed one share the same $\eta$, one
obtains
\begin{equation}
\eta^{\tau}_{0} \simeq  
\frac { \(( 1- \eta^\tau_{1}\)) \(( 1- \eta \))^{k-1}} 
{\sum^{q}_{\sigma=1} \((1- \eta^\sigma_{1} \))\(( 1- \eta \))^{k-1} } =
\frac { \(( 1- \eta^\tau_{1}\))}
{\sum^{q}_{\sigma=1} \((1- \eta^\sigma_{1} \)) }\ .  
\end{equation} 
We immediately compute the derivatives $\frac{\partial
  \eta^1_0}{\partial \eta^1_1}\big|_{\eta_1^\sigma\equiv\eta}$ and
$\frac{\partial \eta^1_0}{\partial
  \eta^2_1}\big|_{\eta_1^\sigma\equiv\eta}$ that form the entries of
the Jacobian matrix $T$ given in Eq.~(\ref{first_kind}):
\begin{equation}
\frac{\partial \eta^1_0}{\partial
  \eta^1_1}\bigg|_{\eta_1^\sigma\equiv\eta} = -\frac 1{q}, ~~~~~~~ 
\frac{\partial \eta^1_0}{\partial 
  \eta^2_1}\bigg|_{\eta_1^\sigma\equiv\eta} = \frac 1{q(q-1)} 
\end{equation} 
Using eigenvalues~(\ref{jac_eig}), we finally find for $c \gg q$ 
\begin{eqnarray}
\begin{cases}
\lambda_1 =\ \ \ \ 
\((\frac{\partial \eta^1_0}{\partial \eta^1_1}
- \frac{\partial \eta^1_0}{\partial \eta^2_1}\))
\bigg|_{\eta_1^\sigma\equiv\eta} \ \ \ \ \ \simeq - \frac 1{q-1} 
\\ \lambda_2 =
\((\frac{\partial \eta^1_0}{\partial \eta^1_1}
+(q-1) \frac{\partial \eta^1_0}{\partial \eta^2_1}\))
\bigg|_{\eta_1^\sigma\equiv\eta} \simeq \ \ \ \ 0
\end{cases}
\end{eqnarray}
Therefore, the stability criterion reads 
\begin{equation}
 \frac{k}{(q-1)^2}< 1
\end{equation}
such that the instability appears at connectivity greater than
\begin{equation}
c_{SP} = (1-q)^ 2 + 1 \simeq  q^2 -2q + 2 + o(1)\ .
\end{equation}
In fact, this formula gives very good results even at small $q$ for
both Erd\"os-R\'enyi and regular random graphs. The reason why it is
working so well is that we are considering connectivities growing like
$q^2$, so the condition $c \gg q$ is satisfied very fast.

The dominant contributions give always an integer $c_{SP}$. In
particular in the case of regular graphs it is therefore important to
know at least the sign of the next-order term. Numerically we find
this to be positive for arbitrary $q$. Since we define $c_{SP}$ as the
first unstable connectivity, we should therefore write
\begin{equation}
c^{Fixed}_{SP} = (q-1)^2 +2.
\end{equation}
This formula amazingly appears to be exact even for $q=3$. Note that,
due to its $q^2$ dependence, as compared to the order $q \log{q}$ of
the $q$-COL/UNCOL transition, this instability is always located in
the UNCOL phase and is thus irrelevant to the physics of the problem.
It is however useful since it tells us when the SP equations at
$y=\infty$ stop to converge on a single graph. The finite $q$ result
is also extremely well approximated by the $c\gg q$ limit in the case
of fluctuating connectivities, giving
\begin{equation}
c^{ER}_{SP} = 1 + (q-1)^2 + o(1).
\end{equation}

\subsubsection{Instability of TYPE-II}

Finally, we have also to check if a TYPE-II instability exists. To do
so, we start our investigation close to the clustering transition,
i.e. we go back to the scaling
\begin{equation}
  \label{eq:kd_asymp_stab2}
  k = q [ \ln q + \ln \ln q + \alpha]\ .
\end{equation}
As shown before, this results in 
\begin{equation}
  \label{eq:eta_stab2_asymp}
  \eta = \frac 1q \left[ 1- \frac \gamma{\ln q} + o(\ln q ^{-1})
  \right] 
\end{equation}
where $\gamma$ is the smaller of the two solutions of $2\gamma
e^{-\gamma} = e^{-\alpha}$. The stability criterion
(\ref{Crit:Fac_II}) reads
\begin{equation}
  \label{eq:stab2}
  \lambda_{TYPE-II} = k(q-1) \frac
  {\sum_{l=0}^{q-2} (-1)^l {q-2 \choose l} [1-(l+2)\eta]^{k-1} }
  {\sum_{l=0}^{q-1} (-1)^l {q \choose l+1} [1-(l+1)\eta]^k }
  <1 \ .
\end{equation}
Plugging in Eqs. (\ref{eq:kd_asymp_stab2}) and
(\ref{eq:eta_stab2_asymp}), and keeping only the leading order terms,
we find
\begin{equation}
  \label{eq:l_stab2}
  \lambda_{TYPE-II} = e^{\gamma - \alpha} + {\cal O}\left( \frac
    {\ln\ln q}{\ln q}\right)  
\end{equation}
For sufficiently large $q$, this becomes smaller than one if and only
if $e^{\gamma - \alpha}<1$ which, according to the condition for
$\gamma$, holds for all $\alpha>1/2$. We therefore conclude that 1RSB
is stable for all connectivities larger than
\begin{equation}
  \label{eq:asymp_stab2}
  c_m = q\ \left[ \ln q + \ln \ln q + \frac 12 +o(1) \right]\ .
\end{equation}
This value is slightly larger than the one of $c_d$, i.e. there is a
linearly growing gap between the onset of a non-trivial 1RSB solution
and its stability. Note, however, that the relative distance of both
expressions is decreasing, the two leading terms in $c_d$ and $c_m$
coincide.

It is very interesting to compare this behavior with the best lower
bounds obtained from heuristic algorithms. As discussed in more detail
in the introduction, these have the leading term $q \ln q$. This
results supports the intuitive feeling that local linear-time
algorithms are not able to enter the clustered phase due to the
proliferation of metastable states of non-zero energy. It would be
interesting to understand better the influence of the structure of the
energy landscape on the behavior of local algorithms, because this
interplay between static and dynamic features may be a useful tool in
systematically improving algorithms.

Note in this context also that the clustered phase covers half of the
COL phase for large $q$. This means that, compared to, e.g., $q=3$,
the influence of the clustering phenomenon becomes asymptotically more
and more important. In this sense, a full understanding of the
clustering phenomenon beyond the 1RSB approximation is of crucial
interest.

\section{The finite energy phase diagram}
\label{Sec:Finite}
\begin{figure}
 \begin{center}
\includegraphics[width=8cm]{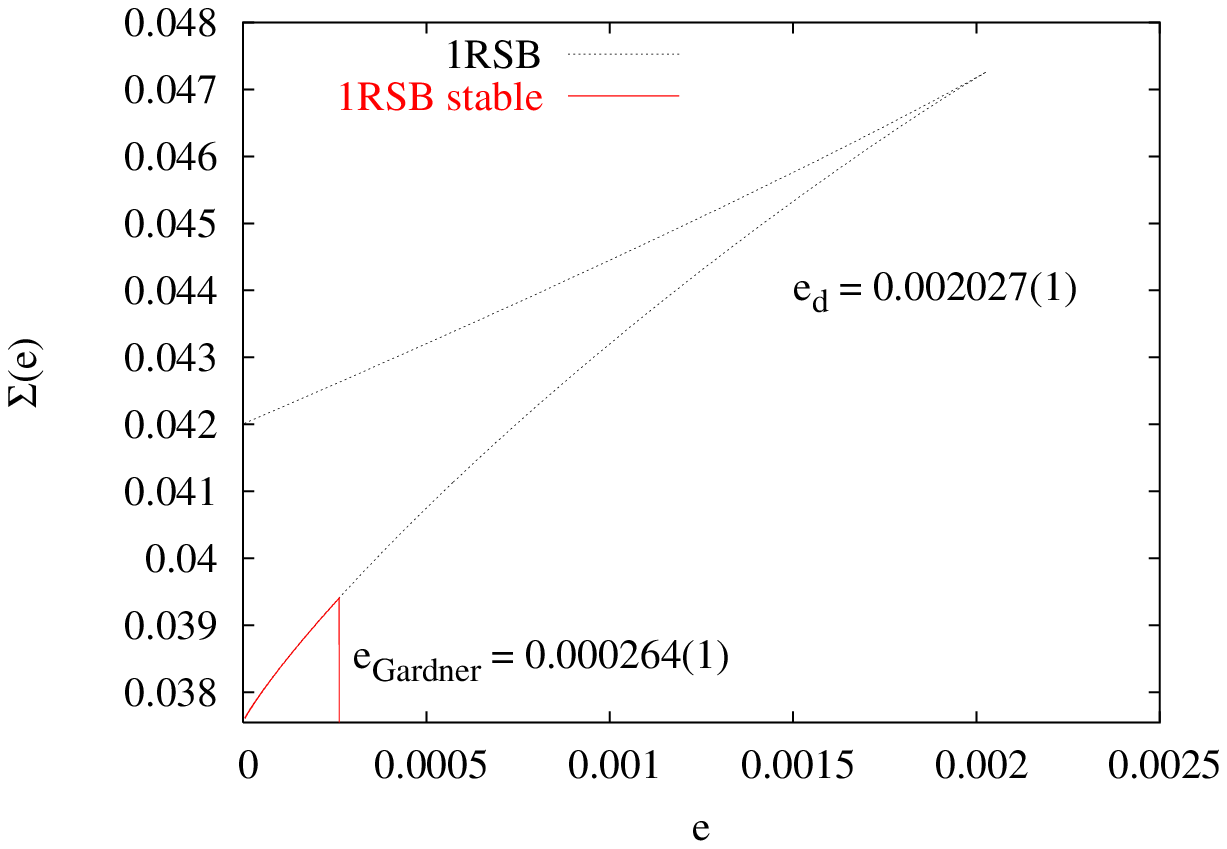}
\includegraphics[width=8cm]{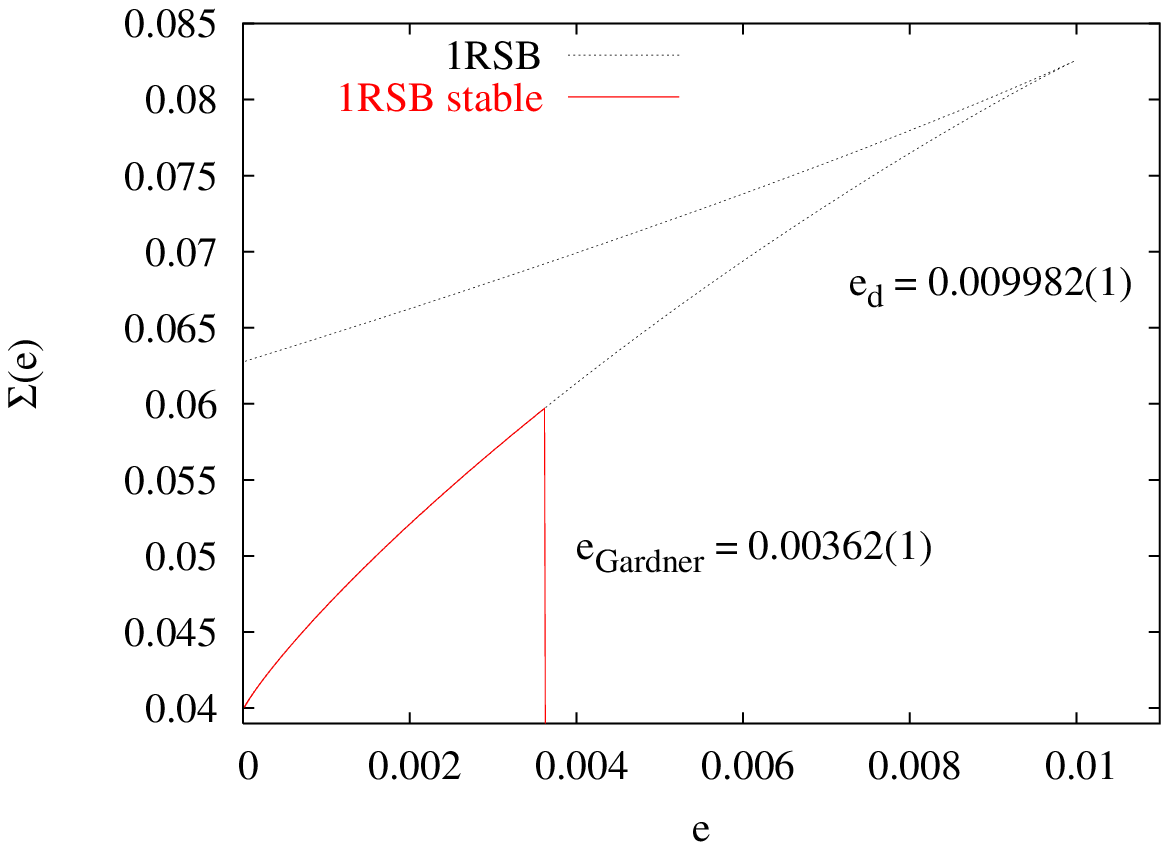}
\end{center}
 \caption{Complexity 1RSB and its 1RSB stable portion for the
 $3$-coloring problem on graph with fixed connectivity $c=5$ (left),
 an for the the $4$-coloring at $c=9$. Both system are in the COL
 Phase, and ground state, as well as low energy states, are described
 by a 1RSB ansatz. However, at high energy, when $e>e_m$, excited
 states should be described by a different ansatz, probably F-RSB.}
 \label{Pic_c_fixed} \end{figure}

It is very interesting to consider also finite-energy states, both for
physical and computer science motivations. Indeed, the nature of these
states and their degree of RSB provide some crucial information for
the physical picture of the model and for its finite-temperature phase
transition. Thermodynamically, the anti-ferromagnetic Potts model on
random graphs behaves in fact like a Potts
glass~\cite{PottsGlass,BOOKS_SG} and it is thus interesting {\it per
  se}. For instance, one would like to know if, depending on the
number of colors and on the connectivity, the system behaves like a
$p-$spin model and has a 1RSB transition, or like the
Sherrington-Kirkpatrick model which has a continuous transition. From
the point of view of combinatorial optimization, it is also widely
believed that the time dependence of local search algorithm, and the
performance they may reach, is also related to the structure of the
energy landscape~\cite{Dyn,Monasson,Gel}. It has been shown very
recently that it is possible to predict a threshold where a slow
annealing will end~\cite{MoRi2} using the statistical physics
information on the nature of excited states.

The basic question we will answer in this section follows
from~\cite{Coloring2}, where the complexity $\Sigma(e)$ where computed
using the 1RSB solution. We know now that the 1RSB solution is in fact
not always correct, so we would like to know which part of the
complexity curve is correct and which part is not.

\subsection{Regular random graphs}
 
Let us first consider regular graphs. In Figs. \ref{Pic_c_fixed}, we
plot complexity versus energy, i.e. the logarithm of the number of
states, divided by $N$, versus their energy, for 3-COL on 5-regular
graphs, and for 4-COL on 9-regular graphs. Both cases are in the COL
phase which implies $\Sigma(e=0) \geq 0$. We already showed in this
paper that the 1RSB approach is stable for these two cases at $e=0$,
and therefore the computation of $\Sigma(e=0)$ is correct.

Extending our work from $e=0$ to positive energies, we indeed find
that only the TYPE-II instability is relevant in the physical part of
the phase diagram (i.e. where complexity is non-negative). Using
Eq.~(\ref{fluctu_stab_II}) at finite $y$, we observe that a large part
of the positive-energy solution is 1RSB unstable. i.e. only a small
fraction of the complexity curve close to the ground state is exact.
Therefore, using the notation $e_G$ (for Gardner
energy~\cite{Gardner}) for the highest energy for which 1RSB is
stable, the 1RSB computation of the complexity is only valid for
energies $0<e<e_{G}$, and should be modified for $e_G<e<e_{d}$.

Two questions are arising. First, how to compute the correct
complexity when 1RSB is unstable? Of course, it is in principle
possible to use a 2RSB ansatz, but this is technically much more
demanding, and again the stability towards 3RSB should be tested.  It
is still a matter of debate how to modify the complexity when 1RSB is
unstable: The usual paradigm is that, if not described by 1RSB, then
states should be described by a full RSB ansatz (since all known 1RSB
unstable models seem to be described by full RSB). Assuming that this
is indeed the case, we have to face the question of the complexity of
full RSB states, which again is still very controversial.  While some
authors argue that full RSB states have a vanishing complexities, the
question remains unsolved and has recently seen a renewed interest
(see~\cite{Complexity}).

Another very important question, using these complexity curves, would
be to predict dynamical thresholds, like in~\cite{MoRi2}.  That would
be a very natural extension of this work to dynamical studies.

\subsection{Erd\"os-R\'enyi random graphs}
\label{Sec:Erdos}

\begin{figure}
\begin{center}
\includegraphics[width=10cm]{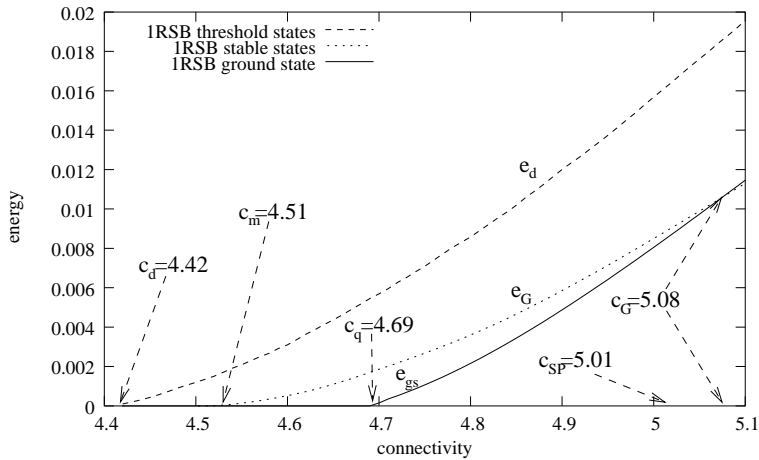}
\end{center}
\caption{Finite energy phase diagram of the $3$-coloring. $e_{gs}$ gives
  the ground state energy density, $e_{G}$ the energy density of the
  1RSB stable states of highest energy, and and $e_{d}$ the threshold
  energy in 1RSB approximation, i.e. the energy below which a
  1RSB solution exists.}
\label{Pic_c3_fluc} 
\end{figure}

For Erd\"os-R\'enyi random graph, we concentrate on $3$-COL. Instead
of showing further complexity curves, we prefer to display, in fig.
\ref{Pic_c3_fluc}, a phase diagram which we think contains a good
summary of this work. The following points are to be commented:
\begin{itemize}
\item First, in full line, we display the ground state energy $e_{gs}$
  versus the average connectivity $c$. It is zero below $c_q$, in the
  COL phase, and positive when $c>c_q \simeq 4.69$, i.e. in the UNCOL
  phase. This has been calculated using the energies for which the
  complexity is exactly zero.
\item The energy $e_{d}$ is the threshold energy, i.e.~the energy
  below which the clustering transition appears, and where replica
  symmetry is broken. In 1RSB approximation, it is non zero for
  $c>c_d\simeq 4.42$.
\item The energy $e_{G}$ is the Gardner energy, which tell us where
  the clustering phenomenon becomes 1RSB unstable. For $e_{gs}<e<e_G$,
  states are 1RSB stable, whereas they are unstable for $e_G<e<e_d$.
  This energy line starts at $c_m$ (note that $c_d<c_m<c_q$) and
  allows us to determine in which zone of this phase diagram 1RSB is
  correct. It crosses the ground state energy line at $c_G \simeq
  5.08$ (we use again the terminology of~\cite{MoRi}; $c_G$ stand for
  Gardner connectivity). Therefore, for $c>c_G$, the problem is never
  1RSB (at least for physical, i.e. positive-complexity solutions).
\item Going back to the $e=0$ line, we have finally, in the UNCOL
  phase, the point where the SP equations stop to converge on a single
  graph. This happens for all $c \geq c_{SP}\simeq 5.01$.
\end{itemize}

\section{conclusion} 
\label{Sec:Concl}
                      
In this paper we have studied in detail the limits of 1RSB
approximation for the $q$-coloring problem on random graph of both
fixed and Poissonian degrees. The 1RSB approximation shows two
different kind of instabilities: a TYPE-I instability where the
clusters of solutions tend to aggregate, and a TYPE-II instability
where solutions inside a cluster form further levels of
clusterization. In the colorable phase, we find in particular the
latter kind of instability. Luckily enough, at any $q\geq 3$, the
COL/UNCOL transition is found in the 1RSB stable region, making it
plausible that the values presented in
Tabs.~\ref{tab:results_fixed}~and~\ref{tab:results_complete} are in
fact the true $q$-COL/UNCOL thresholds.

In the limit of $q\rightarrow \infty$ many simplifications can be done
in the calculation, allowing for a fully analytic treatment of the
problem, and also for checking our results against the rigorous ones
presented in \cite{Lu,Ach_Nao}. We find the COL/UNCOL threshold to be
asymptotically $c_q = 2q \log{q} - \log{q} -1 + o(1)$ which on the one
hand coincides precisely with the rigorous upper bound of {\L}uczac
\cite{Lu}, which on the other hand differs only one from the lower
bound of Achlioptas and Naor \cite{Ach_Nao}. All these findings are
good news for the 1RSB cavity approach, that not only turns out to be
consistent with independently established rigorous mathematical
results, but also allows for sharper, though not rigorous,
determination of threshold values.

There are several lines in which the work presented here could be
extended. The most straight-forward direction is probably the question
of how to implement a 2RSB calculation, in order to understand the
phase-space structure in the 1RSB unstable region, and to see in how
far the 1RSB approximations for the clustering transition, the
complexity and the threshold energy have to be changed. 

A second interesting direction concerns the connection between the
failure of linear-time algorithms and the onset of clustering. Even if
a connection between both seems pretty intuitive due to the existence
of exponentially many metastable states, one should keep in mind that
algorithms do not follow a physical dynamics with detailed balance
etc. The connection between the energy landscape and the
configurations explored by the algorithm is therefore far from beeing
obvious. Even if so far no local linear-time algorithms were found
that solve $q$-COL inside the 1RSB stable region, their mathematical
analysis has no obvious connection to the landscape properties of the
model. It would therefore be extremely interesting to either establish
this connection or to prove its non-existence.

\begin{acknowledgments}
  We thank M.~M\'ezard, F.~Ricci-Tersenghi, O.~Rivoire and R.~Zecchina
  for useful and cheerful discussions. We also thank the hospitality
  of the ICTP Trieste, and two of us (FK and MW) thank the hospitality
  of the ISI in Torino, where part of this work was done.  We
  acknowledge support from the ISI Foundation, the EXYSTENCE Network,
  and from European Community's Human Potential program under
  contracts HPRN-CT-2002-00319 (STIPCO) and HPRN-CT-2002-00307
  (DYGLAGEMEM).
\end{acknowledgments}

\appendix

\section{Instability of the second Kind}
\label{appendix:TypeII}
In this appendix we study the eigenvalues of the matrix $V$ defined
by Eq.~(\ref{eq:V}), 
\begin{equation}
  V_{\tau\to\tilde\tau,\sigma\to\tilde\sigma} = \frac
  {\partial\pi^{\tau\to\tilde\tau}_0 }{\partial
    \pi^{\sigma\to\tilde\sigma}_1 }.
\end{equation}
To compute this matrix element from Eq.~(\ref{eq:type2}) one needs to
consider all configurations of messages $(\sigma_2,...,\sigma_d)$ on
the incoming links $2,3,..,d$ such that\\[-0.66cm]
\begin{itemize}
\item[{\it (a)}] if the incoming warning on link 1 is given by
  $\sigma$, the
  warning $\tau$ is induced at the output, and\\[-0.66cm]
\item[{\it (b)}] if the incoming warning on link 1 is given by
  $\tilde\sigma$, the warning $\tilde\tau$ is induced at the
  output.\\[-0.66cm]
\end{itemize}
Since we have $q+1$ different messages labelled by $0,...,q$, we have
to deal with a $\((q(q+1)\))\times\((q(q+1)\))$ matrix.

A physical interpretation of recursion (\ref{eq:u}) is that, when a
site is exposed to a $q$-component field $\vec h=(h^1,...,h^q)$
(entries are all negative or zero, reflecting the anti-ferromagnetic
nature of the Hamiltonian) then\\[-0.66cm]
\begin{itemize}
\item[{\it (i)}] if there is a unique maximal field component
  $h^\tau$, the site will send a message $-\vec e_\tau$ saying ``Do
  not take color $\tau$'' via the outgoing link (in the following we
  call this a warning of color $\tau$), and\\[-0.66cm]
\item[{\it (ii)}] it will send a zero message otherwise.\\[-0.66cm]
\end{itemize}
Let us study all elements, case by case, first when a zero message is 
changed to a messenger with a color, then a color is
changed into another one.

\subsection{Changing $0$ to a colored message (or a color to $0$)}
When one changes one input message $0$ to a colored one, e.g. $-\vec
e_1$, it can be shown that most corresponding matrix elements vanish.
A simple way to show it is the following: Consider link one, where we
are changing the incoming message, and the other $d-1$ incoming links.
In the first configuration, link one is sending a $0$, so it has no
effect. The other $d-1$ incoming messages induce a field $\vec h$. If
we turn now the message $0$ on branch one to $-\vec e_1$, this has the
effect of decreasing field component $h^1$ by one. The outgoing
message is non-zero only if there is a unique maximum field, this
change has {\it no} effect at all if $h^1$ was not a maximum field
component.  To obtain non-zero matrix elements, we thus have to
consider only situations in which this $h^1$ was a maximum field
before changing the first message.
 
Two cases are arising: 
\begin{itemize}
\item[1)] First, if $h^1$ was the {\it only} maximum component before
  the change, then the output was $e_1$ in the first configuration.
  Adding a new incoming message of color one decreases $h^1$ by one
  unit. Now either $h^1$ is still the unique maximum, and nothing has
  changed, or it now equals to another field, say $h^2$, and the new
  outgoing message becomes $0$. There is thus a finite probability to
  change the input from color $1$ to $0$ by changing an input $0$ to
  color $1$.
\item[2)] Second, if $h^1$ was not the maximum field and if they were
  two such maximum fields, say $h^1$ and $h^2$, then adding an
  incoming message of color $1$ forces $h^1$ to decrease, and
  therefore $h^2$ may become the only maximum and the new output is
  $-\vec e_2$. 
\end{itemize}
There are thus only two non zero terms, that we will call ${\cal{A}}$
and ${\cal{B}}$:
\begin{equation}
\label{def_A_preli}
\begin{cases}
{\cal{A}} = V_{0 \rightarrow 2,0 \rightarrow 1}\\
{\cal{B}} = V_{1 \rightarrow 0,0 \rightarrow 1}
\end{cases}
\end{equation}
All other non-zero terms having a zero input changed to a colored one,
result from simple color permutations and equal the two described
ones. Note that similar equations can be written when one changes an
input of color $1$ to $0$. In this case the only non-zero terms are
given by 

\be
\label{def_A_zero}
\begin{cases}
{\cal{A'}} = V_{2 \rightarrow 0,1 \rightarrow 0}\\ 
{\cal{B'}} = V_{0 \rightarrow 1,1 \rightarrow 0}\\ 
\end{cases}
\ee

The values of ${\cal{A}},{\cal{A'}},{\cal{B}}$ and ${\cal{B'}}$ are
associated to the two following situations (considering the set of
fields $\vec h$ resulting from the other $d-1$ incoming messages): (a)
${\cal{A}}(q,\{{\eta\}})$ and ${\cal{A'}}(q,\{{\eta\}})$ are
probabilities of having $h^1$ and $h^2$ as the two {\it only} maximum
field and (b) ${\cal{B}}(q,\{{\eta\}})$ and ${\cal{B'}}(q,\{{\eta\}})$
are the probability of having $h^1$ as the maximum field, and that
there exists at least one other field with value $h=h^1-1$. To obtain
the final matrix element one has to put the reweighting factor
corresponding the to second configuration, therefore we find:
\bea
\label{expression_abc}
{\cal{A}}(q,\{{\eta\}}) = {\cal{A'}}(q,\{{\eta\}})&=& C_0
\sum_{(h^3,..,h^q > h^1=h^2)} \tilde P(h^1,h^2,h^3,..,h^q) e^{yh^2}\\
\nonumber {\cal{B'}}(q,\{{\eta\}}) &=& C_0 \sum_{\substack{
(h^2,..,h^q > h^1)\\ (h^1-1=max(h^2,h^3,..,h^q))}} \tilde
P(h^1,h^2,h^3,..,h^q) e^{yh^1}\\ \nonumber {\cal{B}}(q,\{{\eta\}}) &=&
C_0 \sum_{\substack{ (h^2,..,h^q > h^1)\\
(h^1-1=max(h^2,h^3,..,h^q))}} \tilde P(h^1,h^2,h^3,..,h^q)
e^{y(h^1-1)} 
\eea 
where we have introduced, following~\cite{Coloring2}, the notation
$\tilde P(\vec{h})$ as the probability of having a configuration of
messages that gives a set of fields $\vec{h}$ before any reweighting
is done. Thus $\tilde P(\vec h)$ would result from
Eq.~(\ref{eq_surv_P}) by setting $y=0$. This notation will be of great
use in the computation of the appendix. Note that, to compute $\tilde
P(\vec h)$ we consider here only messages arriving from the $d-1$
unchanged links $2,3,...d$.
 
Finally, another very important property for the structure of the
matrix is that, by changing $0$ to $\tilde\sigma$, one cannot change a
color $\tau$ to another color $\tilde\tau$ in the output. This will
turn out to considerably simplify the problem.

\subsection{Changing one color to another color}

We now consider a change of the first incoming message from color $1$
to color $2$. One can show, using similar kind of reasoning, that
$V_{1 \rightarrow 2,1 \rightarrow 2}=V_{3 \rightarrow 2,1 \rightarrow
  2}=V_{1 \rightarrow 3,1 \rightarrow 2} = V_{2 \rightarrow 3,1
  \rightarrow 2}=V_{3 \rightarrow 1,1 \rightarrow 2}=0$ and that $V_{2
  \rightarrow 1,1 \rightarrow 2}={\cal{A}}$. Having in mind these
relations, we can now write the stability matrix.

\subsection{The stability matrix $V$}

We write the stability matrix $V$ in a way that justifies why
we did not care about some terms in the previous section. In the base
we chose, the matrix is {\it block triangular}, so we will need only
to care of diagonal block matrices to compute eigenvalues (since, in
a triangular block matrix, eigenvalues are eigenvalues of the
diagonal matrices). We write
\begin{eqnarray}
V= \left[ \begin{array}{c||c|c|}
 &(0,e)~~(e,0)&(e,e')\\
\hline
\hline
\begin{array}{c}
~~~~~~q\begin{cases} (0,e_1) \\ (0,e_2) \\ \dots \\ (0,e_q) \\ \end{cases} \\
~~~~~~q\begin{cases} (e_1,0) \\ (e_2,0) \\ \dots \\ (e_q,0) \\ \end{cases}
\end{array} & M & W \\
\hline
q(q-1)\begin{cases} (e_1,e_2) \\ (e_2,e_3) \\ \dots \\ (e_3,e_2) \\
(e_2,e_1) \\\end{cases} & 0 & Z \\
\end{array}
\right]
\end{eqnarray}
where $M$ and $Z$ can be written
\begin{eqnarray}
M  = \left[ \begin{array}{cccc|cccc}
0 & {\cal{A}} & ... & {\cal{A}} & {\cal{B'}} & 0 & ... & 0 \\
{\cal{A}} & 0 & ... & {\cal{A}} & 0 & {\cal{B'}} & ... & 0 \\
... & ... & ... & ... & ... & ...& ... & ... \\
{\cal{A}} &  {\cal{A}} & ... & 0 & 0 & 0 & ... & {\cal{B'}} \\
\hline
{\cal{B}} & 0 & ... & 0 & 0 & {\cal{A}} & ... & {\cal{A}} \\
0 & {\cal{B}} & ... & 0 & {\cal{A}} & 0 & ... & {\cal{A}} \\
... & ... & ... & ... & ... & ... & ... & ... \\
0   &  0  & ... & {\cal{B}} & {\cal{A}} & {\cal{A}} & ... & 0 \\
\end{array}
\right]
~
\text{and}
~
Z  = \left[ \begin{array}{cccc}
0 & ... & 0 & {\cal{A}} \\
0 & ... & {\cal{A}} & 0 \\
... & ... & ... &... \\
{\cal{A}} & ... & 0& 0 \\
\end{array}
\right]\nonumber
\end{eqnarray}

\subsection{Eigenvalue analysis}

We are now ready to find the biggest eigenvalues of the matrix $V$,
which will be the biggest of all eigenvalues from $Z$ and $M$. The
matrix $Z$ has $\frac{q^2-q}{2}$ eigenvalues $-{\cal{A}}$, and
$\frac{q^2-q}{2}$ eigenvalues ${\cal{A}}$. Eigenvalues of $M$ can
be easily studied using again its block matrix structure; the biggest
one is found to be $(q-1){\cal{A}}
+\sqrt{{\cal{B}}{\cal{B'}}}$. Thus the biggest eigenvalue of
the whole stability matrix is
\begin{equation}
\lambda = (q-1){\cal{A}}(q,\{{\eta\}}) +
\sqrt{{\cal{B}}(q,\{{\eta\}}){\cal{B'}}(q,\{{\eta\}})}
\end{equation}
where $A,B$ and $B'$ are defined by Eq.~(\ref{expression_abc}).
In the $y=\infty$ case, new simplification arises because ${\cal{B}}$
is associated with contradicting messages and it is annihilated by its
reweighting factor. Thus
\begin{equation}
\lambda_{y \rightarrow \infty} = (q-1){\cal{A}}_{y=\infty}(q,\{{\eta\}}) 
\end{equation}
where ${\cal{A}}_{y=\infty}(q,\{{\eta\}})$ may be explicitly computed (see
Appendix~\ref{appendix:TypeII_A}) and reads
\begin{equation}
{\cal{A}}_{y=\infty}(q,\{{\eta\}}) = \frac
{\sum_{l=0}^{q-2} (-1)^l {q-2 \choose l} \prod_{i=2}^k [1-(l+2)\eta_i] }
{\sum_{l=0}^{q-1} (-1)^l {q \choose l+1} \prod_{i=1}^k
[1-(l+1)\eta_i] }
\end{equation}

\section{Instability of the second kind at $y=\infty$}
\label{appendix:TypeII_A}

Let us concentrate on the {\it colorable phase}, where the ground
states are characterized by $y=\infty$. Here we give an explicit
computation of ${\cal{A}}_{y=\infty}$. Let us concentrate for the
derivation for $3$-COL. Following the notation of
App.~\ref{appendix:TypeII}, we denote $\tilde P(\vec h)$ the
probability of having a configuration of $d$ warnings that sum up to
the field $\vec h$ without reweighting. It has no direct physical
meaning but it is of great technical help in the present computation,
cf.~\cite{Coloring2}.

We first need to calculate the value of the normalization constant
$C_0$ in the $y\to\infty$ limit. Since reweighting is killing any
term with positive energy shift, the only surviving terms in the
recursion are those where all fields have at least one zero component,
allowing for the selecting of at least one color without violating an
edge. The normalization factor thus reads
\begin{equation}
\label{eq_C_norm}
\frac 1{C_0} = \tilde P(0,0,0) + 3 \sum_{h^1 < 0} \tilde P(h^1,0,0) + 
3 \sum_{h^1,h^2 < 0} \tilde P(h^1,h^2,0)
\end{equation}
where the combinatorial factors 3 appearing in the r.h.s. are obtained
by noting that $\tilde P(h,0,0)=\tilde P(0,h,0)=\tilde P(0,0,h)$ and
that $\tilde P(h^1,h^2,0)=\tilde P(h^1,0,h^2)=\tilde P(0,h^1,h^2)$.

Now, we need to compute the expression for ${\cal{A}}$ from
Eq.~(\ref{expression_abc}), summing this time only over the $d-1$
incoming warnings $2,3...,d$ in the computation of $\tilde P$ in the
numerator
\be 
\label{cala_1}
{\cal{A}}(q,\{{\eta\}}) = C_0
\sum_{(h^3,..,h^q <h^1=h^2=0)} \tilde P(h^1=0,h^2=0,h^3,..,h^q) 
\ee

Therefore, ${\cal{A}}(q,\{{\eta\}})$ is easily computed once the
expression of $\tilde P$ for a given number of $d$ branches are known.
Using the cavity recursion equations, one can show that, when summing
over $d$ neighbors
\begin{eqnarray}
\tilde P(0,0,0) &=& \prod_{i=1}^k (1-3\eta_i) 
\\
\sum_{h^3<0} \tilde P(h^3, 0, 0 ) &=& \prod_{i=1}^k ( 1 - 2 \eta_i ) -
\tilde P(0,0,0) =  \prod_{i=1}^k ( 1 - 2 \eta_i ) - 
\prod_{i=1}^k (1-3\eta_i)
\\
\sum _{h^1,h^2 < 0} \tilde P(h^1,h^2,0) &=& \prod_{i=1}^k (1 - \eta_i) 
- 2 \sum_{h^1<0} \tilde P(h^1, 0, 0 ) - \tilde P(0,0,0)  \nonumber\\ 
&=&  \prod_{i=1}^k (1-\eta_i) - 2  
\prod_{i=1}^k (1-2\eta_i) + \prod_{i=1}^k (1 - 3\eta_i)
\end{eqnarray}

Using now these relations, with the proper product over the $d$ or
$d-1$ incoming messages, we immediately get from Eq.~(\ref{cala_1})
\begin{equation}
{\cal{A}}_{y=\infty}(q=3,\{{\eta\}}) = \frac
{\prod_{i=2}^k (1-2\eta_i) - \prod_{i=2}^k (1 - 3\eta_i)}
{3 \prod_{i=1}^k (1 - \eta_i)  - 3 \prod_{i=1}^k (1 - 2
\eta_i) +  \prod_{i=1}^k (1 - 3 \eta_i)}.
\end{equation}
This equation can be easily generalized to an arbitrary
number $q$ of colors, and we find
\begin{equation}
{\cal{A}}_{y=\infty}(q,\{{\eta\}}) = \frac
{\sum_{l=0}^{q-2} (-1)^l {q-2 \choose l} \prod_{i=2}^k [1-(l+2)\eta_i] }
{\sum_{l=0}^{q-1} (-1)^l {q \choose l+1} \prod_{i=1}^k
[1-(l+1)\eta_i] }
\end{equation}


\begin{thebibliography}{99}
\bibitem{Ac_note} Private communication with D. Achlioptas and
  C. Moore: Using a pretty physical symmetry assumption in the the
  second moment method of \cite{Ach_Nao}, C. Moore found the
  conjectured lower bound. A complete proof seems, however, very hard
  and is still missing.
\bibitem{note_1rsb} Note that the local stability of the 1RSB solution
  is necessary, but not sufficient for demonstrating its correctness.
  Up to now, however, no model was found which shows a discontinuous
  transition from 1RSB to a more-step solution. Based on this
  experience we conjecture a locally stable 1RSB solution to be in
  fact the exact solution.
\end{thebibliography}
\end{document}